\documentclass[aps,prb,twocolumn,amsmath,showpacs,amssymb,superscriptaddress]{revtex4}

\usepackage{amsmath}
\usepackage{graphicx}
\usepackage{amssymb}
\usepackage{dsfont}
\usepackage{color}

\begin{document}
\title{
Chiral currents in one-dimensional fractional quantum Hall states
}
\author{Eyal Cornfeld}
\affiliation{Raymond and Beverly Sackler School of Physics and Astronomy, Tel-Aviv University, Tel Aviv, 69978, Israel}
\author{Eran Sela}
\affiliation{Raymond and Beverly Sackler School of Physics and Astronomy, Tel-Aviv University, Tel Aviv, 69978, Israel}

\begin{abstract}
We study bosonic and fermionic quantum two-leg ladders with orbital magnetic flux.
In such systems, the ratio, $\nu$, of particle density to magnetic flux shapes the phase-space, as in quantum Hall effects. In fermionic (bosonic) ladders, when $\nu$ equals one over an odd (even) integer, Laughlin fractional quantum Hall (FQH) states are stabilized for sufficiently long ranged repulsive interactions.
As a signature of these fractional states, we find a unique dependence of the chiral currents on particle density and on magnetic flux.
This dependence is characterized by the fractional filling factor $\nu$, and forms a stringent test for the realization of FQH states in ladders, using either numerical simulations or future ultracold-atom experiments.
The two-leg model is equivalent to a single spinful chain with spin-orbit interactions and a Zeeman magnetic field, and results can thus be directly borrowed from one model to the other.
\end{abstract}
\maketitle

\section{Introduction}
\begin{figure}[t!]	
	\centering	
	\includegraphics*[width=.95\columnwidth]{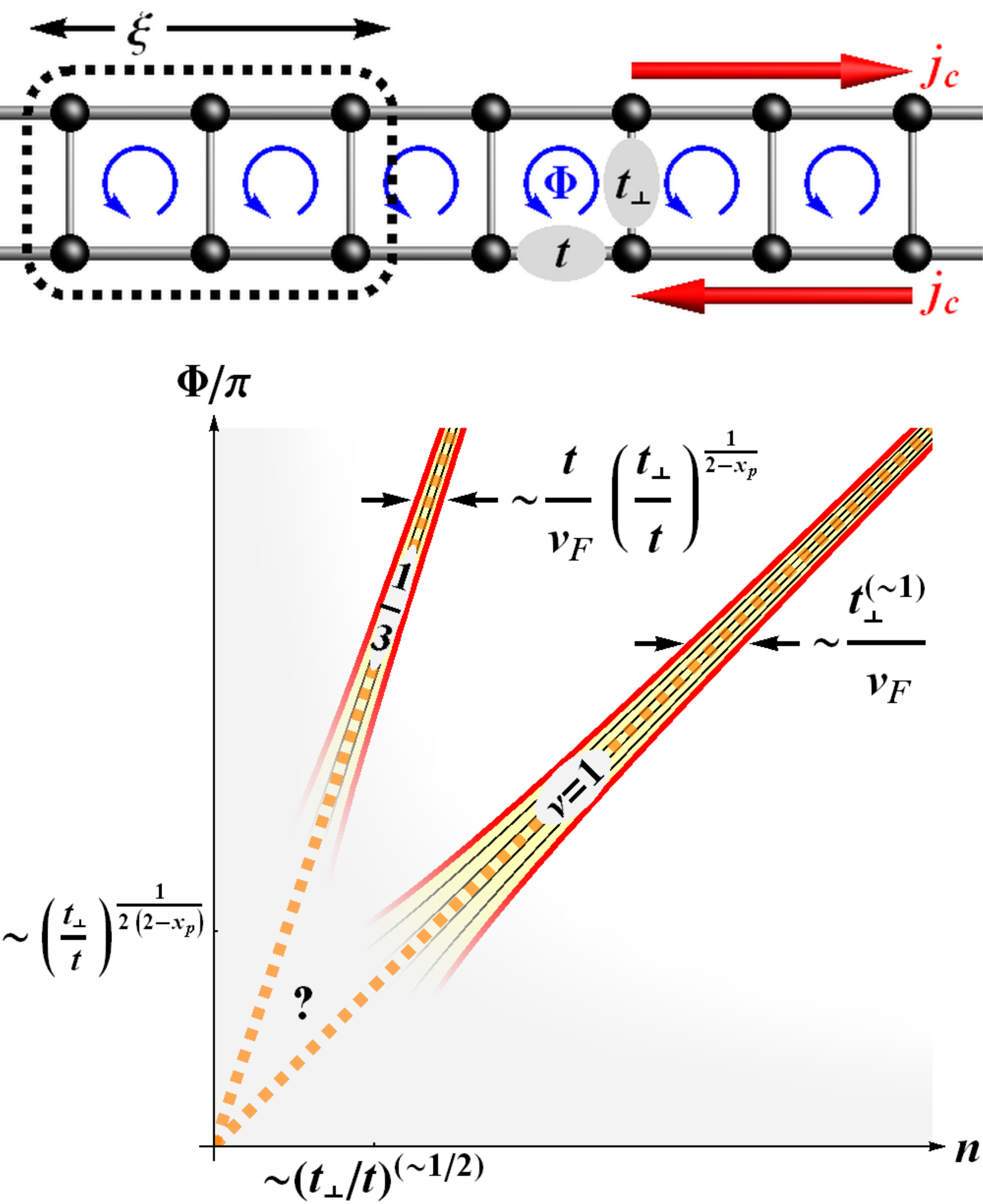}
	\caption{Top: Two-leg ladder model
with magnetic flux per plaquette $\Phi$ leading to chiral currents $j_c$ flowing around the ladder. $\xi$ denotes interaction range. Bottom: Schematic fermionic chiral current contours of Eq.~(\ref{mainres}) marked by thin lines in the $n$-$\Phi$ plane, within an integer and a fractional QH phase. The phase transition lines out of the QH states are marked by solid thick lines, and the thick dotted lines correspond to fillings ${n = \nu\Phi/\pi}$. The validity regime is discussed in the text.}
	\label{fig:ladder}
\end{figure}

Can the quantized Hall effect be observed in one dimension (1D)? Whereas a single 1D chain does not
allow for any orbital magnetic field effects, a ladder system as shown in Fig.~\ref{fig:ladder} is the
minimal extension where these are permitted.
Recently, a realization of bosonic ladders was reported by Atala \emph{et~al.}~\cite{Bloch14} using
ultracold-atoms exposed to a uniform artificial magnetic
field created by laser-assisted tunnelling~\cite{Aidelsburger,Goldman13,Zoller03,Gerbier10,Kolovsky}. They reported an observation of the chiral currents flowing around the ladder due to the effective magnetic field~\cite{Bloch14,hugel}.
Motivated by this experimental ability, the coupled wire realization  of the bosonic Laughling ${\nu=1/2}$ fractional quantum Hall effect (FQHE) introduced by Kane \emph{et~al.}~\cite{kane2002},  was recently suggested in two-leg ladders for strong on-site interactions~\cite{Petrescu,Grusdt}.

An equivalent 1D setup is a single chain with spinful particles; spin-orbit interactions play the role of an orbital magnetic field, and a noncommuting Zeeman field acts as inter-chain hopping. This has been extensively discussed in semiconductor quantum wires with strong Rashba spin-orbit interactions, specifically in the context of helical liquids~\cite{oreg,lutchin}; when the system is strongly interacting, the possibility of a ``fractional helical liquid" was suggested~\cite{OregSelaStern}. As an alternative to spin-orbit interactions in electronic systems, effective spin-orbit coupling and Zeeman field may also be generated in systems of ultracold-atoms confined to 1D~\cite{Lin,Cheuk,Wang,Cui}. Recently, an observation of chiral edge states was achieved using fermionic~\cite{Mancini} and bosonic~\cite{Stuhl} 1D gases with an extra synthetic dimension originating from nuclear spin degrees of freedom.
This setup was theoretically envisioned to stabilise exotic states such as the fractional helical state, and numerically studied using density matrix renormalization group (DMRG) methods~\cite{Fazio,Zeng}.

The aim of this paper is (i) to analytically establish realizations of FQH phases in concrete 1D lattice models and (ii) propose  a physical quantity that unambiguously signals these phases' fractional quantum Hall nature. By FQH in 1D, we refer to the coupled wire construction introduced by Kane \emph{et~al.}~\cite{kane2002}. While the 2D limit of this construction corresponds to the robust quantum Hall liquid, we herein study the thin-stripe regime, in which the system is sensitive to small perturbations, and hence finding detectable signatures is more demanding.

In Sec.~\ref{sec:nonint}, we introduce the instructive noninteracting integer model; in Sec.~\ref{sec:Model}, we elaborate on the realizations of fractional quantum Hall states in 1D two-leg ladders of either interacting bosons or interacting fermions. These FQH instabilities occur when the 1D density $n$ is related to the flux per plaquette $\Phi$ as
\begin{equation}
\label{eq:1}
n=\nu \frac{\Phi}{\pi},
\end{equation}
with ${\nu=1/m}$, and where $m$ is either odd for fermions or even for bosons; see dotted lines in Fig.~\ref{fig:ladder}. The FQH phases support finite deviations from this density, as schematically shown in Fig.~\ref{fig:ladder}, which increase with the inter-chain coupling $t_\perp$ and represent the finite compressibility of the system analogous to the 2D edge states.
Upon increasing the range of interactions, we find that arbitrary Laughlin ${\nu=1/m}$ states can be stabilized even at small inter-chain coupling. The required range of interactions increases~\cite{Fazio} for low filling factors;
the ${\nu=1/3}$ state already requires interactions between nearest neighbour rungs, while the ${\nu=1/2}$ bosonic state is stabilized for sufficiently strong but only on-rung interactions~\cite{Petrescu}.
Interestingly, in the synthetic dimension realizations of quantum ladders, the interactions become non-local in the synthetic dimension~\cite{Mancini,Stuhl,Majoranacoldatoms} (along the rungs), allowing one to reach this bosonic FQH state.

In Sec.~\ref{sec:persistent}, we address the question: What are manifestations of the FQHE in 1D ladders? In Refs.~\onlinecite{OregSelaStern,Meng14,Cornfeld15} transport was considered through leads connected to the 1D fractional helical state. In contrast, we herein wish to discuss thermodynamic bulk observables which may be detected in cold-atom experiments and simple simulations, where transport can not be directly measured. Such an observable is the chiral current $j_c$ that flows in the ground state due to the magnetic flux; see Fig.~\ref{fig:ladder}.
The possibility that chiral currents screen the orbital magnetic field in a kind of a Meissner phase, was pointed out~\cite{Orignac,Petrescu13,Piraud} and recently observed experimentally~\cite{Bloch14,Mueller14}; here we focus on the FQH phase.

We show that within the FQH phase the current depends on density $n$ and flux $\Phi$ as
\begin{equation}\label{mainres}
 j_c \propto \left( n-\frac{\nu}{\pi} \Phi \right),
\end{equation}
up corrections that vanish for small inter-chain coupling $t_\perp$ and are discussed in detail in Sec.~\ref{sec:persistent}. Therefore, for small $t_\perp$, the behaviour described by Eq.~(\ref{mainres})
can be detected in the current map of the $n$-$\Phi$ plane as shown in Fig.~\ref{fig:ladder}; contours of constant current are asymptotically parallel to the constant filling factor line, Eq.~(\ref{eq:1}), which is depicted by the dotted line in Fig.~\ref{fig:ladder}.

These current contour are determined by the fractional filling $\nu$ and hence allows one to measure it. As we show, this result implies the emergence of fractional ``edge" states on the two-leg ladder and thus forms a stringent test for the stabilization of FQH states in ladders.
The constraints on the validity regime of this result are discussed in Secs. \ref{sec:instabilities} and \ref{correction}. Under these constraints the phase diagram in Fig.~\ref{fig:ladder} is exhaustive and no additional phases occur.

We conclude in Sec.~\ref{sec:discuss} providing perspectives on the experimental relevance in ultracold-atomic systems.

\section{Noninteracting model}
\label{sec:nonint}
We consider a spinless two-leg ladder of either fermions or bosons as shown in Fig.~\ref{fig:ladder}. Our main attention in this paper is devoted to the combined effect of interactions and magnetic flux $\Phi$. It is instructive, however, to first display the simple physics of the noninteracting fermionic model, and investigate the properties of the chiral currents in this simpler case, as done in this section.

\begin{figure}
        \centering
        \includegraphics[width=.95\columnwidth]{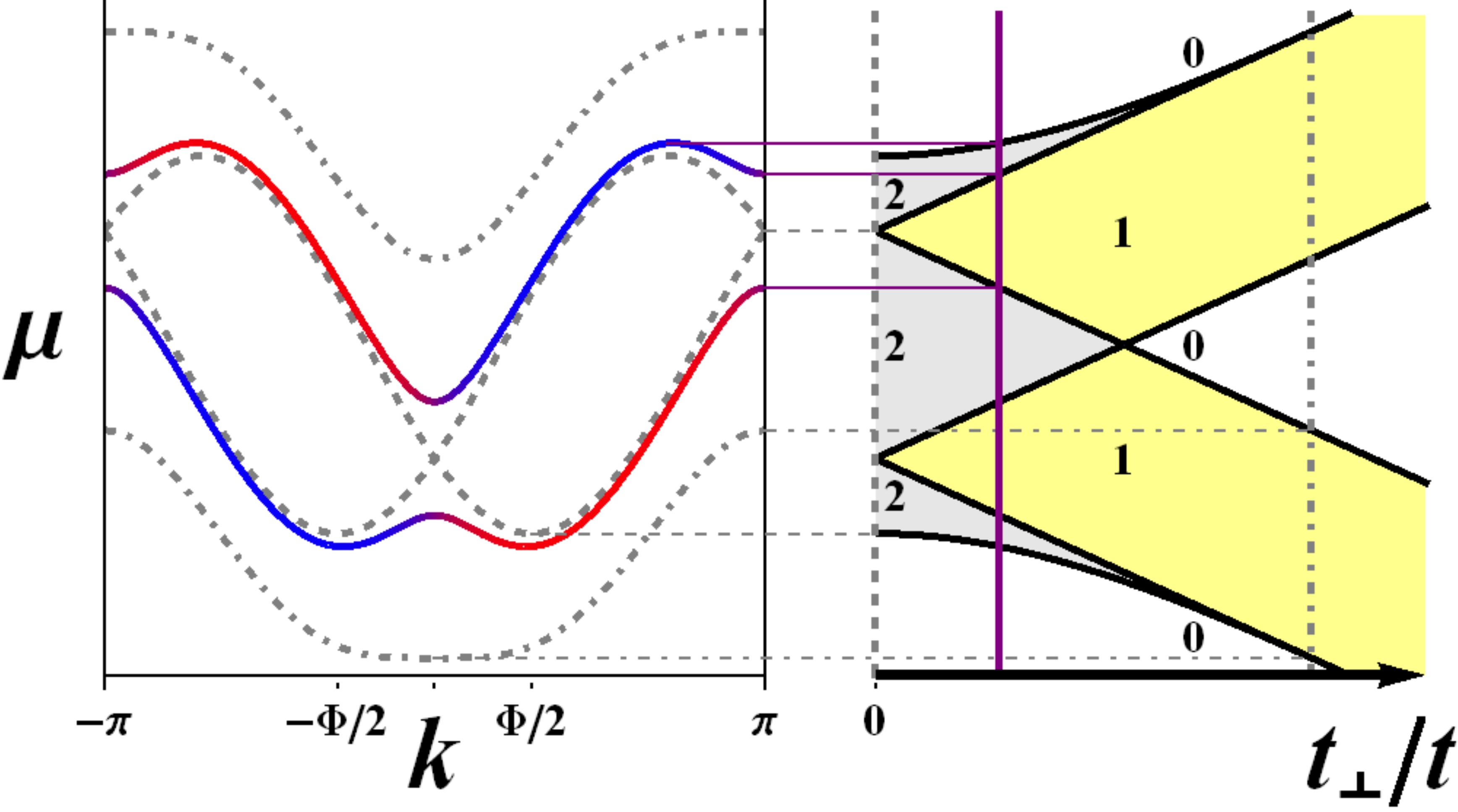}
        \caption{Left: dispersion relation $\epsilon_k^\pm$ for representative values of $t_\perp/t$.
         Right: phase diagram for noninteracting fermions; ${c=0,1,2}$ labels the number of pairs of Fermi points (central charge).}\label{fig:phase-nu-1}
\end{figure}

The noninteracting model Hamiltonian is
$H = H_0+H_{\perp}$, where
\begin{equation}
\label{nonintmodel}
H_0 = -t \sum_{j,y} \left(c^\dagger_{j,y} c^{\phantom{\dagger}}_{j+1,y}+\mathrm{h.c.}\right)
\end{equation}
describes hopping within each chain ${y=1,2}$, and the inter-chain hopping is
\begin{equation}
\label{nonintHperp}
H_{\perp} = - t_\perp \sum_j \left(c^\dagger_{j,1} c^{\phantom{\dagger}}_{j,2} e^{i \Phi j}+ \mathrm{h.c.}\right).
\end{equation}
Here $\Phi$ is the magnetic flux per plaquette.
It is convenient to make the gauge transformation ${c'_{j1} = c_{j1} e^{- i\frac{1}{2} \Phi j}}$, ${c'_{j2} = c_{j2} e^{ i \frac{1}{2}\Phi j}}$, which moves the phase factor $e^{i \Phi j}$ from inter- to intra-chain hopping, yielding the Bloch Hamiltonian
 \begin{equation}
 \label{eq:Bloch}
 H_k = \left(
\begin{array}{cc}
-2t\cos(k-\frac{\Phi}{2}) & t_{\perp} \\
t_\perp & -2t\cos(k+\frac{\Phi}{2}) \\
\end{array}
\right)
 \end{equation}
(the lattice constant is set to unity).
Its eigenvalues $\epsilon_k^\pm$ are plotted in Fig.~\ref{fig:phase-nu-1} for various values of $t_\perp/t$. At ${t_\perp=0}$ we have two cosine dispersions shifted horizontally by ${\pm \frac{\Phi}{2}}$ (dashed lines). Any small inter-chain coupling $t_{\perp}$ opens a gap at the crossing points (full lines). As seen in the right panel of Fig.~\ref{fig:phase-nu-1}, upon scanning the chemical potential one finds two partially gapped ``chiral" regions with only one pair (${c=1}$) [rather than two pairs (${c=2}$)] of Fermi points, where $c$ denotes the central charge. In these chiral phases, the left- and right-moving modes reside on distinct chains; see color code in Fig.~\ref{fig:phase-nu-1}. Upon further increasing $t_\perp$, the chiral nature of the 1D modes is gradually reduced, and one eventually obtains (dot-dashed) dispersion curves which are nonoverlapping bands dominated by the inter-chain hopping.

We shall focus on the regime of ${t_\perp \ll t}$ near the instability leading to the opening of the lower partially gapped ${c=1}$ region in the the phase diagram.
The particle density is defined as ${n = \sum_{y=1,2}  \langle c^\dagger_{j,y} c^{\phantom{\dagger}}_{j,y} \rangle }$, and the gap opens at the Fermi level when ${n=\frac{\Phi}{\pi}}$. Borrowing the 2D definition of filling factor, namely the ratio of particle density \emph{per site}, ${\langle  c^\dagger_{j,y} c^{\phantom{\dagger}}_{j,y} \rangle =\frac{n}{2}}$, to the density of flux quanta per plaquette,
\begin{equation}
\label{fillingF}
\nu =  \frac{\pi n}{\Phi},
\end{equation}
we see that the gap opening at small $t_\perp$ occurs at unit filling factor ${\nu=1}$. This gap opening is just the two wire version of the wire-construction of the quantum Hall effect~\cite{kane2002}. One pair of modes is gapped out, and one pair remains gapless in analogy with the chiral edge states in the integer quantum Hall effect.

The two-leg ladder is equivalent to a Rashba wire upon reinterpretation of (i) the two legs of the ladder as a spin degree of freedom, (ii) the inter-chain hopping $t_\perp$ as a spin-flipping Zeeman field, and (iii) the magnetic flux $\Phi$ as a Rashba spin orbit coupling causing a momentum shift ${\pm k_{SO}}$ of the two dispersions,
\begin{equation}
\begin{array}{ccc}
y=1,2 & \leftrightarrow & \sigma = \uparrow, \downarrow, \\
t_\perp & \leftrightarrow & B_\mathrm{Zeeman}, \\
\Phi & \leftrightarrow & 2 k_\mathrm{SO}.
\end{array}
\end{equation}
After this relabeling of indices and parameters, the partially gapped chiral state with ${c=1}$ corresponds to a helical state with the two spins propagating to opposite directions~\cite{oreg,lutchin}. In this context, $j_c$ is simply a persistent spin current~\cite{Yi06}.
We next discuss the persistent current $j_c$ generated by the magnetic flux in the two-leg ladder.

\subsection{Chiral current in the $n$-$\Phi$ plane}
\label{nonintphin}
The magnetic flux induces a persistent current in the ground state (GS), see Fig.~\ref{fig:ladder},
\begin{equation}
\label{jaPhi}
j_c=-\left\langle\frac{\partial H}{\partial\Phi_\mathrm{tot}}\right\rangle_\mathrm{GS} = - \frac{1}{L}\frac{\partial E_\mathrm{GS}}{\partial\Phi}  ,
\end{equation}
where ${\Phi_\mathrm{tot} = L \Phi}$ is the total flux in a two-leg ladder of length $L$. Here, we expressed the current as the derivative of the ground state energy ${E_\mathrm{GS} = \langle H \rangle_\mathrm{GS} }$ with respect to $\Phi$. We shall see that the current can be used to probe 1D quantum Hall physics.

\begin{figure}
\centering
\includegraphics*[width=.95\columnwidth]{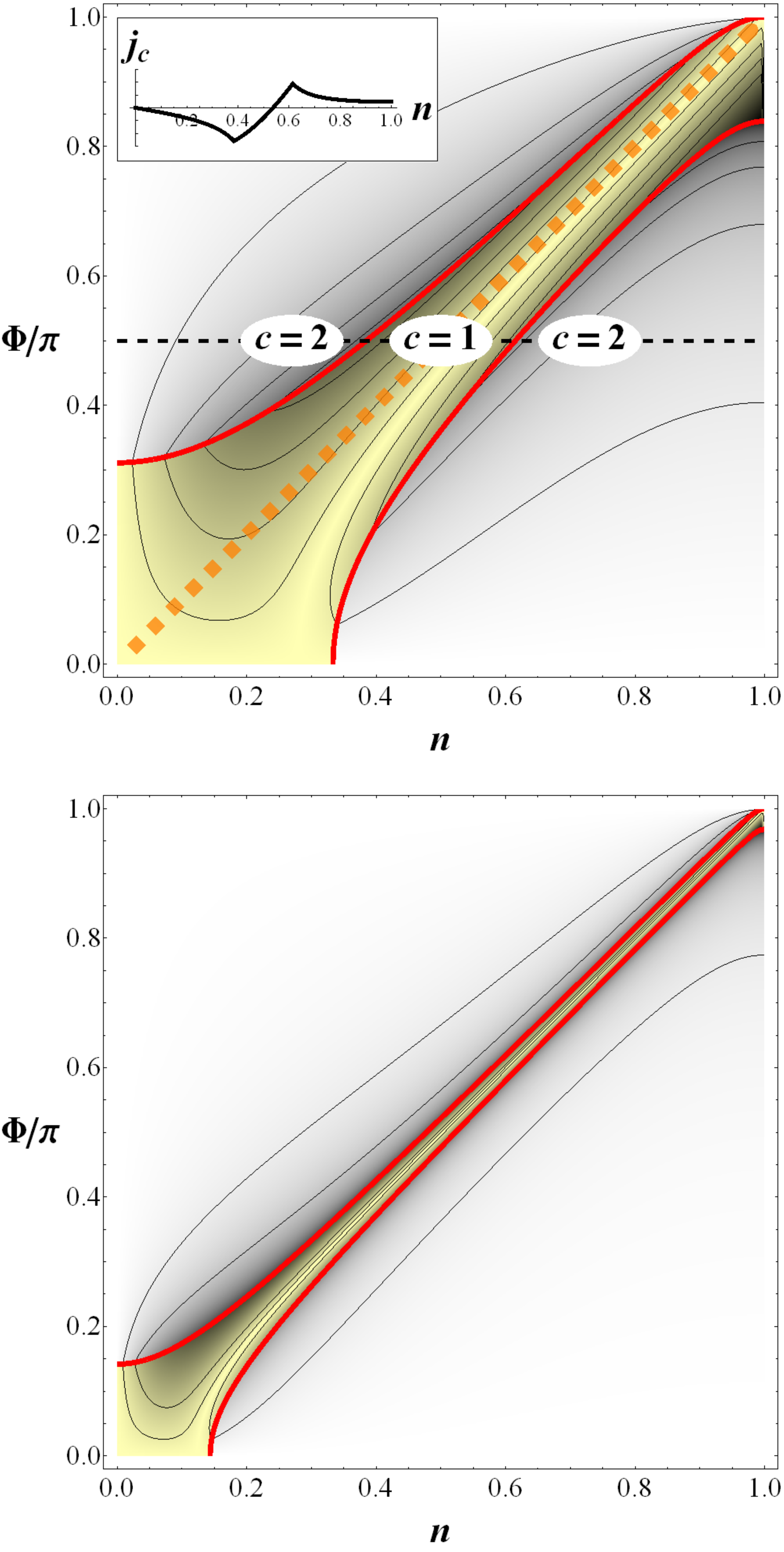}
\caption{Contours of the chiral current $j_c$ marked by thin lines in the $\Phi$-$n$ plane for ${t_\perp =0.5 t}$ (top) and ${t_\perp=0.1 t}$ (bottom). The phase transition lines are marked by solid thick lines, and the thick dotted line corresponds to ${n = \Phi/\pi}$. The inset (top) depicts the chiral current along the horizontal dashed line.}
\label{fig:nphi}
\end{figure}

One may explore the current dependence on density $n$ and flux $\Phi$. In a noninteracting model the ground state energy is the sum over individual occupied states. In the ${c=1}$ regime only one band contributes,
\begin{equation}
\label{eqdjdmnon}
j_c(n,\Phi)=-\int_{- \pi n}^{\pi n} \frac{dk}{2\pi}\frac{\partial\epsilon_k^-}{\partial\Phi}.
\end{equation}
The contours of ${j_c(n,\Phi)}$ are plotted in Fig.~\ref{fig:nphi} for two values of ${t_\perp/t}$ both in and out of the chiral ${c=1}$ phase. The phase transitions are marked as thick red lines, and the current has cusp singularities at these transitions, as shown in the inset. We see that within the ${c=1}$ chiral phase and for not too small a density, the current contours are approximately parallel to the line defined in Eq.~(\ref{fillingF}). This behaviour becomes more pronounced as ${t_\perp/t}$ becomes smaller; see lower panel of Fig.~\ref{fig:nphi}.

For a quantitative description of this assertion we decompose the symmetric integral Eq.~(\ref{eqdjdmnon}) into contributions from the Fermi sea and from near the Fermi surface (${n\simeq\frac{1}{\pi}\Phi}$), as ${j_c=j_c^\mathrm{sea}+j_c^\mathrm{surf}}$, where
\begin{align}
\label{separate}
j_c^\mathrm{sea}=-\int_{0}^{\Phi}\frac{dk}{\pi}\frac{\partial\epsilon_k^-}{\partial\Phi},~~~
j_c^\mathrm{surf}=-\int_{\Phi}^{\pi n}\frac{dk}{\pi}\frac{\partial\epsilon_k^-}{\partial\Phi}.
\end{align}
Relegating details to Appendix~\ref{appendixNONINTj}, using this decomposition we find that the total current satisfies
\begin{equation}
\label{relationnonint}
(\partial_n+\pi\partial_\Phi)j_c=A^\mathrm{surf} \cdot (n-\tfrac{\Phi}{\pi})+ A^\mathrm{sea} \cdot \tfrac{t_\perp^2}{t^2}\ln\tfrac{t}{t_\perp}+\mathcal{O}(\tfrac{t_\perp^2}{t^2}).
\end{equation}
The two terms in the right hand side arise from the Fermi surface and the Fermi sea, their coefficients $A^\mathrm{surf}$ and $A^\mathrm{sea}$ are derived in Appendix~\ref{appendixNONINTj}. The Fermi surface contribution originates from the band curvature and is given by  ${A^\mathrm{surf}=\tfrac{1}{4\pi}\partial_n^2\mu}$.

As herein explained, two terms on the right hand side of Eq.~(\ref{relationnonint}) form a small correction.
If one tunes the density to the exact filling factor ${\nu=1}$, the first correction term vanishes. Deviations of the density ${\delta n}$ within the ${c=1}$ phase are of the order of the energy  gap ${E_\mathrm{gap}=\epsilon_{k=0}^+ - \epsilon_{k=0}^-= 2 t_\perp}$ times the density of states $\frac{1}{\pi v_F}$, where $v_F$ is the Fermi velocity. The first correction term  thus approaches values of order ${\mathcal{O}(t_\perp/t)}$ near the boundaries of the ${c=1}$ region. As a consequence, this leading correction, as well as the second term and other subleading corrections, are altogether negligible for small inter-chain coupling, ${t_\perp\ll t}$, resulting in ${(\partial_n+\pi\partial_\Phi)j_c \simeq 0}$. It implies that indeed, contours of the current are nearly parallel to the dashed line ${n = \Phi/\pi}$ as seen in Fig.~\ref{fig:nphi} for gradually decreasing $t_\perp$.

This behaviour is valid for density ${\sqrt{t_\perp /t} \ll n \ll 1}$. For too small a density, one reaches the situation where the Fermi energy ${\thicksim t n^2}$ becomes smaller than the energy gap ${\thicksim t_\perp}$, giving the lower bound. For too high a density of order unity, the partial gaps of electrons and holes (see Fig.~\ref{fig:phase-nu-1}) approach each other and cause undesired lattice effects. Thus, for ${t_\perp /t \ll 1}$ we have a large region in the parameter space where the contours are asymptotically parallel to ${n = \Phi/\pi}$.
This is a general feature which holds true for the interacting fractional case, as we find in Sec.~\ref{sec:persistent}.

\section{Fractional states in lattice models with long range interactions}
\label{sec:Model}
We now consider either interacting fermions or bosons on the two-leg ladder, and add interactions to the lattice model,
${H = H_0+H_\perp+ H_{\mathrm{int}}}$, with
\begin{equation}
\label{intmodel}
H_{\mathrm{int}} = \sum_{r \ge 0} V(r) \sum_j n_j n_{j+r}.
\end{equation}
Here, ${c^{\phantom{\dagger}}_{j,y} c^\dagger_{j',y'} -(+)  c^\dagger_{j',y'} c^{\phantom{\dagger}}_{j,y}=\delta_{y,y'} \delta_{j,j'}}$ for bosons (fermions) and, ${n_j =  \sum_{y=1,2} c^\dagger_{j,y} c^{\phantom{\dagger}}_{j,y}}$.
In this model, the interaction potential is independent of the rungs' indices $y$ and $y'$, and depends solely on the linear distance $r$ via ${V(r)}$, specified below. Similar to the Hubbard model, the dependence of the interaction only on the total density ${\sum_{y} c^\dagger_{j,y}c^{\phantom{\dagger}}_{j,y}}$ makes the model ${H_0+H_{\mathrm{int}}}$ to be ${\mathrm{SU}(2)}$ invariant with respect to rotations of the spinor ${(c_{j,1}, c_{j,2})}$.

To treat these interactions, we utilize the Luttinger liquid theory. For a generic interaction $H_{\mathrm{int}}$ and at ${t_{\perp}=0}$, the long wavelength behaviour of the quantum ladder can be described by a two component Luttinger liquid (LL)~\cite{Giamarchi}.
The free part of the Hamiltonian is expressible in terms of the density operators of the two chains. It is convenient to introduce bosonic fields $\phi_\mu$, so that the long wavelength fluctuations of the total charge $(\rho)$ and the relative density ($\sigma$), denoted ``charge" and ``spin", respectively, are represented by
\begin{align}
\label{eq:density}
  c^\dagger_{j,1} c^{\phantom{\dagger}}_{j,1} +  c^\dagger_{j,2} c^{\phantom{\dagger}}_{j,2}  &\thicksim n - \frac{\sqrt{2}}{\pi} \nabla \phi_\rho (x),  \hbox{ and}\nonumber \\
 c^\dagger_{j,1} c^{\phantom{\dagger}}_{j,1} -  c^\dagger_{j,2} c^{\phantom{\dagger}}_{j,2} &\thicksim - \frac{\sqrt{2}}{\pi} \nabla \phi_\sigma (x).
\end{align}
The free Hamiltonian can be written as
\begin{align}
\label{HLL}
\mathcal{H}_\mathrm{LL} =\sum_{\mu = \rho,\sigma} \frac{1}{2 \pi} \int dx \left[ v_\mu K_\mu ( \pi \Pi_\mu)^2+ \frac{v_\mu}{K_\mu}  (\nabla \phi_\mu)^2 \right],
\end{align}
where the $\Pi_\mu$ fields are canonically conjugate to the densities, ${[\phi_\mu(x) , \Pi_{\nu}(x')]=i \delta_{\mu \nu} \delta(x-x')}$.
The velocities $v_{\rho,\sigma}$ and LL parameters $K_{\rho,\sigma}$ depend on the strength of interactions and on the density; in the noninteracting fermionic case ${K_{\rho} = K_\sigma=1}$ and ${v_{\rho} = v_\sigma = v_F}$. Note, that for generic interactions, one still has ${K_\sigma=1}$ \emph{if} the model is SU(2) symmetric~\cite{Giamarchi}, as holds true in our model at ${t_\perp=0}$.

The remaining of the section is divided into two parts. We first list the possible perturbations to the Luttinger liquid model~\cite{Orignac,OregSelaStern,Petrescu13}, which include the FQH instability on which we focus. This enables one to obtain conditions for the Luttinger liquid parameter $K_{\rho}$ and density $n$, under which operators identified as opening FQH gaps are most relevant in the renormalization group (RG) sense. We then discuss a specific form of the interaction ${V(r)}$, corresponding to a finite range with hard core interactions, for which $K_{\rho}$ can be computed exactly for ${t_\perp=0}$, assuring that the FQH instability dominates for finite small inter-chain coupling $t_\perp$.

\subsection{Luttinger liquid instabilities}
\label{sec:instabilities}
There are various operators correcting the LL Hamiltonian. We may group them as ${\mathcal{H} = \mathcal{H}_\mathrm{LL} + \delta \mathcal{H}+ \delta \mathcal{H}_\perp}$, where ${\delta \mathcal{H}_\perp}$ includes terms generated by $t_\perp$, while ${\delta \mathcal{H}}$ includes terms which exist at ${t_\perp=0}$. We systematically list these various operators in Appendix~\ref{appendixRELEVANCY}.

The interesting FQH physics would not occur without the inter-chain coupling
\begin{equation}
c^\dagger_{j 1} c^{\phantom{\dagger}}_{j 2}e^{i \Phi j}+\mathrm{h.c.}.
\end{equation}
We may express the particle creation and annihilation operators in the bosonized language,
\begin{equation}\label{eq:psi}
\begin{gathered}
c^\dagger_{j,y} \to \Psi^\dagger_y (x) \sim \sum_p\psi^\dagger_{y,p}(x), \\
 \psi_{y,p}^\dagger(x) = e^{i 2p (\pi \frac{n}{2} x - \phi_y(x))} e^{- i \theta_y(x)} \\
 \phi_{\rho,\sigma} = \tfrac{\phi_{1} \pm \phi_2}{\sqrt{2}} ,~~\theta_{\rho,\sigma} = \tfrac{\theta_{1} \pm \theta_2}{\sqrt{2}},~~~\pi \Pi_\mu = \nabla \theta_\mu(x),
\end{gathered}
\end{equation}
where the sum over $p$  runs over integers for bosons or half integers for fermions~\cite{Giamarchi}.
The operators generated by this expansion are of the form
\begin{equation}
\mathcal{O}_{p} \thicksim \psi^\dagger_{1,-p} \psi^{\phantom{\dagger}}_{2,p}e^{i\Phi x}+\mathrm{h.c.}.
\end{equation}
Such operators may be incorporated into the bosonised LL Hamiltonian using
\begin{equation}
\label{cos}
\mathcal{O}_{p} \thicksim g_p \int dx  \cos (\sqrt{2}\theta_\sigma  - 2p \sqrt{2}\phi_\rho - (\Phi - 2  p \pi n) x),
 \end{equation}
with a coupling constant $g_p$ generated by interactions and by the inter-chain coupling $t_\perp$. These operators are oscillating unless
\begin{align}
\label{eq:simposc}
2p \pi n=  \Phi.
\end{align}
When relevant, such operators may open an energy gap even for a finite deviation from this exact flux up to a commensurate-incommensurate transition~\cite{Orignac,Petrescu13}. The above condition is met when the filling factor Eq.~(\ref{fillingF}) is given by
\begin{equation}
\label{simpnupp}
\nu = \frac{1}{2p},
\end{equation}
which is the inverse of an even(odd) integer for bosons(fermions). The states generated by this operator correspond to the Laughlin FQH phases as constructed by coupled wires~\cite{kane2002}.
The scaling dimension of $\mathcal{O}_{p}$ is given by
\begin{equation}
x_{p} = \frac{1}{2} \left(\frac{1}{K_\sigma} +  (2p)^2 K_\rho \right).
\end{equation}
In the ${\mathrm{SU}(2)}$ symmetric case of ${K_\sigma=1}$, its relevancy ${x_p<2}$ is ensured by a small value of the charge LL parameter
\begin{equation}
\label{Krhocrit}
K_\rho < \frac{3}{(2p)^2}=3 \nu^2.
\end{equation}
In this symmetric case, usual RG analysis~\cite{Giamarchi} shows that the energy gap scales as
\begin{equation}
\label{Egap}
E_\mathrm{gap} \thicksim t (t_\perp/t)^{\frac{1}{2-x_{p}}}.
 \end{equation}
In the noninteracting case for example, the scaling dimension equals ${x_p=1}$ which is consistent with the energy gap ${E_\mathrm{gap} = 2 t_\perp}$.

Additional phases generated by $t_\perp$, described in Ref.~\onlinecite{Petrescu}, are included for completeness in Appendix~\ref{appendixRELEVANCY}, such as the Meissner phase and the vortex lattice~\cite{Orignac}. In order to stabilize a given FQH state, one needs to (i) have a small value of $K_\rho$ satisfying Eq.~(\ref{Krhocrit}), and  (ii) make sure that other instabilities are less relevant. Below we consider a specific interaction ${V(r)}$ with a finite range $\xi$, and  determine the range required to satisfy these conditions and observe a FQH state at any desired fractional filling $\nu$.

\subsection{Solvable Lattice Model}
\label{se:specialmodel}
We now specialize to an interaction potential that vanishes beyond the interaction range $\xi$, see Fig.~\ref{fig:ladder},
\begin{equation}
\label{interaction}
V(r) =
\begin{cases} U v(r)  & \mbox{for } r \le \xi,  \\ 0  & \mbox{for } r > \xi. \end{cases}
\end{equation}
Here, ${v(r)}$ is a positive decreasing function whose specific form is not important, with ${v(0)=1}$ and ${v(r) = \mathcal{O}(1)}$ for ${r \le \xi}$.
The case $\xi=0$ corresponds to on-rung interactions, namely ${V(r=0)}$ comprises both on-site interactions (meaningful for bosons) and interactions between particles on different sites of the same rung (as in the Hubbard model).

We focus on the regime ${U \gg t}$, for general interaction range $\xi$.
The analysis starts by assuming infinite $U$, for which ${H_0+H_\mathrm{int}}$ is exactly solvable, and then relaxes this assumption to finite but large $U$.
In this hard-core limit, the interaction becomes a constraint: states containing two particles horizontally separated by $\xi$ sites or less acquire a very high energy ${\mathcal{O}(U)}$. One can construct a low-energy subspace where the shortest linear inter-particle distance exceeds $\xi$. The allowed states for $N$ bosonic or fermionic particles on the ladder of length $L$ and open boundary conditions are then in one to one correspondence~\cite{SelaPereiraORBITAL,SelaGarst11} with the states of a constrained model. This model consists of $N$ fictitious particles on a ladder of reduced length ${L'=L - (N-1) \xi}$ subject to an additional constraint of not having two fictitious particles on the same rung. Each particle in the reduced lattice corresponds to one particle and $\xi$ empty rungs to its right on the original lattice.

For ${t_\perp=0}$, the leg index ${y=1,2}$ of each particle is conserved \emph{i.e.} the sequence of $y$ values for $N$ particles from left to right ${\{y_1 , y_2 ,...,y_N \}}$ is conserved (considering open boundary conditions for simplicity). For a fixed value of this list, the $\xi-$constrained motion of the $N$ particles, with states labeled by  ${\{ j_1,j_2,...,j_N \}}$ on the ladder, becomes equivalent to free fermions on a single chain of length $L'$; the Pauli principle fully accounts for the interaction constraint.
Following the methods described in Ref.~\onlinecite{SelaGarst11} and detailed in Appendix~\ref{se:appendix1}, we find that the Luttinger liquid parameter describing the model ${H_0+H_{\mathrm{int}}}$ depends on the density per rung $n$ and interaction range $\xi$ as
\begin{equation}
\label{eq:Krho}
K_\rho  \xrightarrow[U \to \infty]{} \frac{1}{2}(1- n \xi)^2.
\end{equation}
For ${\xi=0}$ this result coincides with the exact solution of the Hubbard model \emph{i.e.} ${K_\rho = \frac{1}{2}}$ for infinite repulsion. This result enables us to choose values for $\xi$ and $n$ that yield a sufficiently small $K_\rho$ that satisfies Eq.~(\ref{Krhocrit}). Thus we can use the exact solvability of ${H_0+H_{\mathrm{int}}}$, and proceed using usual RG methods to treat $t_\perp$ as a perturbation, and deduce under which conditions is the FQH cosine perturbation relevant and flows to strong coupling.

The Luttinger parameter ${K_\rho = K_\rho[U/t]}$ is a continuous decreasing function of ${U/t}$, hence having a large enough but finite value of ${U/t}$ leads only to negligible (positive) corrections to the lower bound of $K_\rho$ in Eq.~(\ref{eq:Krho}). At the same time, as is known for the Hubbard model, in the strict ${U  = \infty}$ limit one has a vanishing spin velocity ${v_\sigma \to 0}$, making the Luttinger liquid description  pathological. This is not the case, however, for finite $U$ on which we focus, where the spin velocity remains finite.

In the infinite $U$ limit and for an interaction range $\xi$, the maximal possible density is ${n <  \frac{1}{1+\xi}}$. On the other hand, Eq.~(\ref{eq:Krho}) and the relevancy condition Eq.~(\ref{Krhocrit}) impose a minimal possible density as well,
\begin{equation}
	\label{nrange}
	\frac{1-\sqrt{6}\nu}{\xi}< n< \frac{1}{1+\xi}.
\end{equation}
Comparing the minimal and maximal densities, one finds that the required interaction range for a relevant FQH perturbation at filling factor $\nu$ is
\begin{equation}
\label{range}
\xi \ge \left\lceil \frac{1}{\sqrt{6} \nu}-1 \right\rceil.
\end{equation}
For bosons at ${\nu=1/2}$, this relevancy condition is satisfied for on-rung interactions ${\xi=0}$~\cite{Petrescu13}. However for fermions at ${\nu=1/3}$, or bosons at ${\nu=1/4}$, one needs at least \emph{nearest neighbour} interactions, ${\xi=1}$. This is summarized in Table \ref{tb:xi}.

\begin{table}
 \caption {Conditions for the realization of the 1D FQH states at filling factor $\nu$ in a two-leg ladder with hard-core interaction of range $\xi$ given in Eq.~(\ref{nrange}) and Eq.~(\ref{range}) .}
 \label{tb:xi}
\begin{center}
\begin{tabular}{r || c | c | c | c || l}
  & $\nu=\frac{1}{2}$ & $\nu=\frac{1}{3}$ & $\nu=\frac{1}{4}$ & $\nu=\frac{1}{5}$ &\\ \hline\hline
  $\xi=0$ & $0<n$ & - & - & - & $n<1$ \\ \hline
  $\xi=1$ & $0<n$ & $0.18<n$ & $0.38<n$ & - & $n<0.5$ \\ \hline
  $\xi=2$ & $0<n$ & $0.09<n$ & $0.19<n$ & $0.25<n$ & $n<0.33$ \\ \hline
  $\xi=3$ & $0<n$ & $0.06<n$ & $0.12<n$ & $0.17<n$ & $n<0.25$ \\
\end{tabular}
\end{center}
\end{table}

Note, that in the hard core limit with interaction range $\xi$, the kinetic motion freezes at the maximal allowed density of ${n=\frac{1}{1+\xi}}$. This corresponds to a Mott insulating state.
However, this charge-density wave (CDW) instability is not relevant and hence prevented for  ${n<\frac{1}{1+\xi}}$. Going through the list of operators detailed in Appendix~\ref{appendixRELEVANCY} we find that there are no other nonoscillating relevant operators that compete with the relevant FQH operator Eq.~(\ref{cos}).

\subsubsection{Spin lattice implementation of hard core bosons}
We discuss a simplification and a specific implementation of the bosonic ${\nu=1/2}$ state for on-rung interactions $\xi = 0$. In order to substantially reduce the size of the Hilbert space keeping the essential physics  intact, as we are interested in the limit of large $U$, we can switch from bosons, to a spin-$1/2$ lattice~\cite{Piraud},  using the replacement
\begin{equation}
	c^\dagger_{j,y} \rightarrow S^+_{j,y},~~~~c_{j,y} \rightarrow S_{j,y}^-, ~~~n_j \rightarrow 1+ \sum_{y=1,2} S^z_{j,y}.
\end{equation}
The two-leg ladder model becomes
\begin{align}\label{spins}
H &= -t \sum_{j,y=1,2} \left(S^+_{j,y} S^{-}_{j+1,y}+\mathrm{h.c.}\right)\nonumber \\
&- t_\perp \sum_j \left(S^+_{j,1} S^-_{j,2} e^{i \Phi j}+ \mathrm{h.c.}\right) \nonumber \\
&+ 2 U \sum_j S^z_{j,1} S^z_{j,2}.
\end{align}
Recently, integer Chern insulating phases have been directly discussed in similar XY spin chains~\cite{Grass} with a synthetic magnetic flux.

As a self-consistent example, one may consider a ladder of length ${L \thicksim 100}$ in the vicinity of ${\Phi = 0.8 \pi}$ and ${n=0.4}$ (or equivalently ${\langle S^z \rangle=-0.3}$). There are 2 particles (up spins) every 5 rungs, and hence, given the short range repulsion, the system is far from any CDW instability.  For this setup, ${x_p = 3/2}$ and the gap scales as ${E_{gap} \thicksim t (t_\perp/t)^2}$. The length scale over which the gap is formed scales as
\begin{equation}
\ell^* \thicksim (t/t_\perp)^2.
\end{equation}
The length of the ladder, $L$, should be larger than this crossover scale for the RG flow to fully develop the cosine perturbation from weak to strong coupling. This limits the inter-chain coupling ${t_\perp/}t$ to be not smaller than of order ${\sim 10^{-1}}$.

\hfill

To conclude this section, we have shown that FQH states occur as ground states in two-leg ladders with interactions of sufficiently long range. However, what are signatures of the fractional filling in these ground states? Below, we shall focus on this question.

\section{Chiral current}
\label{sec:persistent}
So far we have discussed explicit lattice realizations of FQH states in 1D ladders. However, when such models are realized, \emph{ e.g.} in an experiment or a numerical simulation, it is not obvious what are their signatures. Here, we explore the chiral current
flowing around the ladder, and find signatures in the $n$-$\Phi$ plane that are characteristic of the fractional filling factor $\nu$, generalizing the behaviour found for the noninteracting case in Sec.~\ref{nonintphin}.

The below calculation of the current and its derivatives with respect to $\Phi$ and $n$ is done using bosonization. It is known that all filled states contribute to the
chiral current, which is a persistent current; the chiral current in the ladder is thus generally \emph{not} an infra-red phenomenon
and cannot be fully accounted
for by an effective low-energy theory~\cite{Narozhny,Carr}. Nevertheless, certain aspects of the persistent current, such as its derivative with respect to particle number or flux, can be computed from the low-energy theory~\cite{Simon01}. Indeed, in Sec.~\ref{nonintphin}, we have identified two contributions to the current in the noninteracting case, one of which is a Fermi surface effect and the other is a Fermi sea effect. When generalizing to the interacting fractional case, the former can be safely extracted from the low-energy theory of bosonization.

We take the following three steps which eventually provide a complete picture of the current in the $n$-$\Phi$ plane: (i) In Sec.~\ref{sec:relation}, we analyse the effects of the cosine perturbation Eq.~(\ref{cos}), which being a relevant operator, flows to strong coupling ${g_p \to \infty}$. It yields straight current contours parallel to the line ${n = \nu \Phi/\pi}$. (ii)
Then, in Sec.~\ref{sec:relationBC}, we include band curvature irrelevant cubic terms, such as  ${(\nabla \phi_\rho)^3}$, in the bosonized Hamiltonian; see Eq.~(\ref{eq:H3}) below. These terms contribute to corrections analogous to the term ${A^\mathrm{surf}\cdot(n- \frac{\Phi}{\pi})}$ in Eq.~(\ref{relationnonint}), and signal deviations of the density from the exact fractional filling.
(iii) Finally, in Sec.~\ref{correction}, we add additional quadratic operators in the LL theory which were not allowed at ${t_\perp=0}$, such as
\begin{equation}
\label{coupling}
\nabla \phi_\rho  \nabla \theta_\sigma,~~~{\mathrm{and}}~~~\nabla \phi_\sigma  \nabla \theta_\rho.
\end{equation}
These corrections give additional distortions of the current contours, which may be nevertheless neglected for small ${(t_\perp/t)^2}$.

The result of these calculations is summarized in the schematic Fig.~\ref{fig:ladder}.

\subsection{Chiral current in FQH states}
\label{sec:relation}
We treat $v_{\rho,\sigma}$ and $K_{\rho,\sigma}$ in Eq.~(\ref{HLL}) as the effective parameters resulting from the RG flow, and treat the strong coupling limit of Eq.~(\ref{cos}). We wish to find a relation between the three susceptibilities
\begin{equation}
\chi_{ab}=\frac{1}{L}\frac{\partial^2 E_\mathrm{GS}}{\partial a \partial b},~~~a,b = n,\Phi .
\end{equation}
Here $\chi_{nn}$ is the charge susceptibility, $\chi_{\Phi \Phi}$ is the diamagnetic susceptibility, and the mixed susceptibility ${\chi_{n  \Phi} = \chi_{\Phi n }}$ describes the change of the persistent current ${j_c = - \frac{1}{L}\partial_\Phi E_\mathrm{GS}}$ with respect to particle addition.

We may opt to treat the fields $\phi_\rho,\theta_\sigma$ as generalized coordinates ($q$) and their canonical conjugates $\nabla\theta_\rho,\nabla\phi_\sigma$ as momenta ($p$). This is a useful choice as the Hamiltonian naturally decomposes to ${\mathcal{H}=\mathcal{H}_p+\mathcal{H}_q}$ as
\begin{align}
\mathcal{H}_p=\frac{1}{2\pi}&\int dx \left[v_\rho K_\rho (\nabla\theta_\rho)^2+ \frac{v_\sigma}{K_\sigma}  (\nabla\phi_\sigma)^2\right], \\
\mathcal{H}_q=\frac{1}{2\pi}&\int dx\left[ \frac{v_\rho}{K_\rho}  (\nabla\phi_\rho)^2+v_\sigma K_\sigma (\nabla\theta_\sigma)^2\right] \nonumber\\
+g_p&\int dx \cos\left(\sqrt{2}\theta_\sigma  - \sqrt{2}\nu^{-1}\phi_\rho +\delta\Phi x\right),
\end{align}
where ${\delta\Phi=\Phi-\tfrac{\pi n}{\nu}}$.
The argument of the cosine is pinned in the strong coupling limit and we thus integrate out the $\theta_\sigma$ field so the Hamiltonian $\mathcal{H}_q$ takes the form
\begin{equation}
\mathcal{H}_q=
\frac{1}{2\pi}\int dx\left[v_\sigma K_\sigma (\tfrac{1}{\nu}\nabla\phi_\rho+\tfrac{1}{\sqrt{2}}\delta\Phi)^2+ \frac{v_\rho}{K_\rho}  (\nabla\phi_\rho)^2\right].
\end{equation}
By recalling Eq.~(\ref{eq:density}) one sees that
a variation of the density is equivalent to setting ${\nabla \phi_\rho = -\frac{\pi}{\sqrt{2}}\delta n}$, and hence
\begin{equation}
\label{EGS}
\tfrac{1}{L}E_\mathrm{GS}\simeq
\frac{\pi}{4}\left[v_\sigma K_\sigma (\tfrac{1}{\nu}\delta n-\tfrac{1}{\pi}\delta\Phi)^2+ \frac{v_\rho}{K_\rho} {\delta n}^2\right].
\end{equation}
Notice that both ``momenta" fields $\theta_\rho,\phi_\sigma$ appear quadratically in $\mathcal{H}_p$ and it may be rigorously integrated out  by choosing the gauge of ${\partial_t\Phi=0}$ in the Lagrangian formalism. This allows us to directly read off the susceptibilities
\begin{gather}
\chi_{n\Phi} = - \frac{1}{2\nu}  v_\sigma K_\sigma ,~~~\chi_{\Phi\Phi} = \frac{1}{2\pi}  v_\sigma K_\sigma, \nonumber\\
\chi_{nn} = \frac{\pi}{2} \left( \frac{v_\rho}{K_\rho}+ \frac{1}{\nu^2}v_\sigma K_\sigma\right).
\end{gather}
The relation ${\partial_n\partial_\Phi E_\mathrm{GS} =-\frac{\pi}{\nu} \partial_\Phi \partial_\Phi E_\mathrm{GS}}$ allows one to extract the fractional filling factor $\nu$ from a ratio of two thermodynamic susceptibilities. Equivalently, from the definition of the chiral current Eq.~(\ref{jaPhi}) we obtain
\begin{equation}
\label{relation}
\left(\partial_n  + \frac{\pi}{\nu} \partial_\Phi \right) j_c=0.
\end{equation}
Notice that this relation does not depend on any of the LL parameters. Below we see that corrections to this relation are only of order ${\mathcal{O}(\delta n ,\delta \Phi )}$ and are thus small in ${t_\perp/t}$. Moreover, the current can be obtained by differentiation of Eq.~(\ref{EGS}),
\begin{equation}
\label{eq29}
j_c = \frac{1}{2 \nu} K_\sigma v_\sigma \left( n -  \frac{\nu}{\pi}\Phi\right).
\end{equation}
These results suggest that in the $n$-$\Phi$ plane the current is constant parallel to the line of fractional filling, ${n =\nu\Phi/\pi}$, within the partially gapped phase; see dotted lines in Fig.~\ref{fig:ladder}.

It is desirable to arrive at a physical interpretation of these results (for this discussion, we retain the electron charge $e$ and Plank constant $\hbar$).

Consider the physics on the edge of an incompressible FQH droplet at filling factor ${\nu=1/m}$. The dynamical degree of freedom is the density of the chiral edge, $n_c$, which is a deformation of the edge of the incompressible liquid~\cite{wen_book}; the chiral current is given by ${j_c = e v n_c}$, where $v$ is the velocity of the edge. Electrodynamics on the edge is quite unusual, as the electron charge is intrinsically entangled with the electromagnetic potential on the edge~\cite{Ezawa}
\begin{equation}
\label{Ezawa}
n_c \to   n_c - \frac{e}{2 \pi m \hbar}A_\parallel.
\end{equation}
It coincides with the minimal substitution argument for the vector potential $A_\parallel$ along the edge. This implies~\cite{Ezawa} the celebrated Laughlin argument of an adiabatic insertion of a flux quantum ${\delta A_\parallel = \frac{2\pi\hbar}{e L_\mathrm{edge}}}$ leading to the total change in the charge of the edge by a fractional amount ${e/m}$ (and creating a quasihole in the flux insertion point). As a consequence, Eq.~(\ref{Ezawa}) demonstrates that the current
\begin{equation}
j_c = e v \left(n_c - \frac{e}{2 \pi m \hbar}A_\parallel\right),
\end{equation}
remains invariant for a simultaneous change of the particle number and of the magnetic flux while keeping their ratio pinned to the filling factor. The two-leg ladder thus can be thought of as an ultra thin FQH droplet, with one chiral edge on the ${y=1}$ sites and the opposite chiral edge on the ${y=2}$ sites. We conclude that the physical meaning of current contours along the ${n = \nu \Phi/\pi}$ lines is the emergence of a fractional chiral edge.
It is important to note that even in the 2D quantum Hall limit, the contours are not exact straight lines. As is discussed below, the main contribution to deviations from linear behaviour stems from band curvature effects, reflecting the dependence of the edge velocity $v$ on the chemical potential, which holds true even in the 2D limit.

\subsection{Band curvature}
\label{sec:relationBC}
At ${t_\perp \to 0}$ where the ${\mathrm{SU}(2)}$ symmetry holds, there are only four cubic terms that may be added to the LL Hamiltonian,
\begin{multline}
\label{eq:H3}\mathcal{H}_3 = \int dx \left[ c_1 \nabla \phi_\rho (\nabla \theta_\sigma)^2 + c_2 (\nabla \phi_\rho)^3\right. \\
\left. + c_3 \nabla \phi_\rho (\nabla \phi_\sigma)^2 + c_4 \nabla \phi_\sigma \nabla \theta_\sigma \nabla \theta_\rho  \right].
\end{multline}
As the other coefficients of cubic terms~\cite{Imambekov,SelaPereira}, $c_1$ satisfies the phenomenological relation
\begin{equation}
\label{c1}
c_1 = - \frac{1}{\sqrt{2}\pi^2} \frac{\partial \left(v_\sigma K_\sigma\right)}{\partial n}.
\end{equation}
When rigorously integrating out these interactions, various polynomial and rational function terms appear in the Hamiltonian. Nevertheless, by following the same procedure and doing some tedious algebra, one obtains the linear correction of order ${\mathcal{O}(\delta n,\delta \Phi)}$ to Eq.~(\ref{relation}),
\begin{equation}\label{relationCORRECTIONbandcurvature}
\left(\partial_n  + \frac{\pi}{\nu} \partial_\Phi \right) j_c=A^\mathrm{surf}\cdot\left(n - \frac{\nu}{\pi}\Phi \right),
\end{equation}
where ${A^\mathrm{surf}= \frac{1}{2 \nu} (\partial_n+\frac{\pi}{\nu} \partial_\Phi) [v_\sigma K_\sigma]}$.
The band-curvature term ${c_1\propto\partial_n[v_\sigma K_\sigma]}$ can be incorporated into a density dependence of the parameter ${v_\sigma K_\sigma}$ in the LL theory Eq.~(\ref{HLL}). The dependence of the current on the density and magnetic flux in Eq.~(\ref{eq29}) thus remains the same up to order ${\mathcal{O}(\delta n,\delta \Phi)^2}$.

The result Eq.~(\ref{relationCORRECTIONbandcurvature}) generalizes the noninteracitng formula Eq.~(\ref{relationnonint}) to the fractional case. As a consistency check, in the noninteracting ${\nu=1}$ case, one has ${K_\sigma=1}$ and ${v_\sigma = \frac{1}{\pi}\partial_n\mu}$, and hence ${A^\mathrm{surf} = -\frac{\pi^2 c_1}{\sqrt{2}} = \frac{1}{4\pi}\partial_n^2\mu}$, which exactly matches the direct results of Sec.~\ref{nonintphin}.

\subsection{Further corrections and validity regime}
\label{correction}
We now consider the terms in Eq.~(\ref{coupling}). It can be shown that they are generated via RG by finite inter-chain hopping ${t_{\perp}\Psi^\dagger_1 \Psi^{\phantom{\dagger}}_2}$ to second order. It is straightforward to include these terms in the analysis of the current by re-evaluating the susceptibilities $\chi_{ab}$, and obtain additional corrections to the right hand side of Eq.~(\ref{relation}) which are quadratic in $t_\perp$. Having already identified the quadratic (and logarithmic) corrections  in the noninteracting case from the \emph{Fermi sea} contribution of all filled states, see Eq.~(\ref{relationnonint}), we deduce that the low-energy theory is not appropriate to evaluate them. Indeed, it misses the logarithmic correction in the term ${A^\mathrm{sea} t_{\perp}^2 \ln t_\perp}$ term in Eq.~(\ref{relationnonint}). Hence, the explicit calculation of the order ${\mathcal{O} \left(t_\perp^2/t^2\right)}$ corrections is superfluous. We conclude that additional corrections to the right hand side of Eq.~(\ref{relationCORRECTIONbandcurvature}) are quadratic in the small parameter  ${t_\perp/t}$ up to logarithmic corrections.

We wish to compare this quadratic correction with the right hand side of Eq.~(\ref{relationCORRECTIONbandcurvature}), ${A^\mathrm{surf} \cdot \left(n - \frac{\nu}{\pi}\Phi \right)}$.  The density deviation, ${\delta n}$, within the partially gapped phase scales as ${E_\mathrm{gap} \propto t_\perp^{\frac{1}{2-x_{p}}}}$ (which for the noninteracting case ${x_p=1}$ behaves like $t_\perp$). We see that as long as the cosine is relevant with scaling dimension ${x_p<3/2}$, the quadratic correction is negligible in the entire FQH phase for small enough ${t_\perp/t}$; this is assumed in Fig.~\ref{fig:ladder}. Otherwise, one may still focus on the vicinity of the exact filling factor and observe lines parallel to ${n = \nu \Phi/\pi}$.

The lowest density for which the linearity of the contours apply is determined by requiring that the kinetic energy ${\thicksim t n^2}$ well exceeds the energy gap. This yields ${n \gg (t_\perp/t)^{\frac{1}{2 (2-x_p)}}}$; see Fig.~\ref{fig:ladder}. On the other hand, the density should be small compared to unity, otherwise, lattice effects would take effect~\cite{Fazio}. Therefore, we expect the behavior in Fig.~\ref{fig:ladder} to apply for ${(t_\perp/t)^{\frac{1}{2 (2-x_p)}} \ll n \ll 1}$.

\section{Discussion}
\label{sec:discuss}
We have studied two-leg ladders of interacting particles with an orbital magnetic field. For sufficiently strong and long ranged interactions, fractional quantum Hall phases are stabilized. Inside these phases there are chiral currents whose contours in the plane of density versus flux are approximately parallel to the line with fractional filling factor, similar to the Landau fan. This behaviour of the current is a signature of the emergence fractional chiral excitations. It distinguishes the 1D FQH state from other phases containing chiral currents such as the Mott insulator phases~\cite{Piraud,Petrescu,Dhar,Wei,Ahmet,Natu}.

Stabilization of FQH phases for small inter-chain coupling and low filling factors ${\nu <1/2}$ requires interaction range beyond on-rung. The cold-atomic technology
(see \emph{e.g.} Refs.~\onlinecite{Bloch14,Mancini})
however involves primarily on-atom interactions. Nevertheless, the on-atom interaction becomes nonlocal in the synthetic dimension~\cite{Mancini,Stuhl,Majoranacoldatoms}. This is sufficient to realize simple bosonic Laughlin states with ${\nu = 1/2}$ even for small inter-chain coupling.

Moreover, many-body systems with tailored long-range interactions have been achieved in Rydberg atoms~\cite{LukinDipoleBlockade,UrbanDipoleBlockade} due to strong van der Waals interaction, yielding Rydberg crystallization~\cite{Weimer10,SelaGarst11} as observed experimentally~\cite{Schauss}.
Arbitrarily small values of $K_\rho$ may be reached in principle for a particle-particle interaction of the form ${V(r) \propto r^{-\beta}}$
for which values of $K_\rho$ have been found analytically~\cite{Dalmonte10}, and where ${\beta=3}$ corresponds to dipolar atoms. Yet, the combination of a synthetic magnetic field and long range interactions, required for the low filling factor FQH states, remains an experimental task.

Nevertheless, powerful numerical techniques~\cite{Fazio,Zeng} have been recently used to simulate the fractional states considered here. With the findings of the current paper it becomes possible to test their fractional quantum Hall nature.

\acknowledgements
We thank M. Becker and S. Trebst for extensive supporting unpublished numerical calculations in the initial stages of this work; E. Dalla Torre, S. Furukawa, M. Goldstein, R. Pereira, and  J. Ruhman for illuminating discussions; and S. Barbarino, R. Fazio, L. Mazza, D. Rossini, and L. Taddia for showing us their work~\cite{Fazio} prior to publication. This work was supported by Israel Science Foundation grant 1243/13 and Marie Curie CIG grant 618188.

\appendix
\numberwithin{table}{section}

\section{Chiral current vs. $n$ and $\Phi$ in the noninteracting case}
\label{appendixNONINTj}
In this appendix we explicitly calculate, to leading orders in $t_\perp$, the contributions to the chiral current from the Fermi surface and the Fermi sea in Eq.~(\ref{relationnonint}).
\subsection{Fermi surface contribution}
Expanding the Fermi surface contribution in Eq.~(\ref{separate}) around small deviations from the ${n=\Phi/\pi}$ filling we get
\begin{equation}
j_c^\mathrm{surf}=-(n-\tfrac{\Phi}{\pi})\partial_\Phi\mu,
\end{equation}
where ${\mu=\epsilon^-_{k=\pi n}(\Phi)}$. We take the derivative along the constant filling factor to get
\begin{equation}
(\partial_n+\pi\partial_\Phi)j_c^\mathrm{surf}=-(n-\tfrac{\Phi}{\pi})(\partial_n+\pi\partial_\Phi)\partial_\Phi\mu.
\end{equation}
We use the property ${2\pi\partial_\Phi\mu+\partial_n\mu=O(t_\perp^2)}$ and relate this expression to the band curvature to leading order in $t_\perp$
\begin{equation}
(\partial_n+\pi\partial_\Phi)j_c^\mathrm{surf}=(n-\tfrac{\Phi}{\pi})A^\mathrm{surf},
\end{equation}
where a direct calculation yields
\begin{equation}
A^\mathrm{surf}\equiv\tfrac{1}{4\pi}\partial_n^2\mu=t\tfrac{\pi}{2}\cos\tfrac{\Phi}{2}+O(t_\perp^2).
\end{equation}

\subsection{Fermi sea contribution}
We turn our attention to the Fermi sea contribution in Eq.~(\ref{separate}). In the limit of ${t_\perp=0}$, the relation ${2\partial_\Phi\epsilon^-_k(\Phi)+\partial_k\epsilon^-_k(\Phi)=0}$ yields
\begin{equation}
\left.j_c^\mathrm{sea}\right|_{t_\perp=0}=-\int_{0}^{\Phi}\frac{dk}{2\pi}\partial_k\epsilon^-_k(\Phi)=-\frac{1}{2\pi}(\epsilon^-_k(\Phi)-\epsilon^-_k(0))=0.
\end{equation}
Hence, by noticing that the main contribution to the integral arises from ${k\ll\Phi}$ we impose a cutoff ${\frac{t_\perp}{t} \ll \Lambda \ll \Phi}$
\begin{equation}
j_c^\mathrm{sea}\simeq-\int_{0}^{\Lambda}\frac{dk}{\pi}\left\{\partial_\Phi\epsilon^-_k(\Phi)-[\partial_\Phi\epsilon^-_k(\Phi)]_{t_\perp=0}\right\}.
\end{equation}
This integral may be directly evaluated in the limit of small ${\Lambda \ll \Phi}$
\begin{gather}
\left.\partial_\Phi\epsilon^-_k(\Phi)\right|_{k\leq\Lambda}\simeq t\cos k \sin\tfrac{\Phi}{2}-\tfrac{t^2 k^2 \sin\Phi}{\sqrt{t_\perp^2+2t^2(1-\cos\Phi)k^2}},\\
\begin{split}
j_c^\mathrm{sea}\simeq t\left[\frac{t_\perp}{t}\right]^2\left\{a_{-1}(\Phi)\ln\frac{t\Lambda}{t_\perp}+\sum_{j=0}^\infty a_j(\Phi)\left[\frac{t_\perp}{t\Lambda}\right]^j\right\}
\\
= t a_{-1}(\Phi)\left[\frac{t_\perp}{t}\right]^2\ln\frac{t}{t_\perp}+O(t_\perp^2).
\end{split}
\end{gather}
We identify ${A^\mathrm{sea}\equiv t\pi\partial_\Phi a_{-1}= t\frac{1+\cos^2\frac{\Phi}{2}}{16\sin^3\frac{\Phi}{2}}}$. This completes the derivation of  Eq.~(\ref{relationnonint}).

\section{Luttinger liquid instabilities for interacting ladders}
\label{appendixRELEVANCY}
In this appendix we review the construction of corrections to the Luttinger liquid Hamiltonian Eq.~(\ref{HLL}), leading to the operators with scaling dimension and commensurability conditions summarized in Table ~\ref{tb:operators}.

Using Eq.~(\ref{eq:psi}), charge conservation in each chain at ${t_{\perp}=0}$ restricts general operators to be of the form
\begin{equation}
 \label{psi4}
\mathcal{O}_{p_1,p_2,p_3,p_4} =(\psi^\dagger_{1,p_1} \psi^{\phantom{\dagger}}_{1,p_2}) (\psi^\dagger_{2,p_3} \psi^{\phantom{\dagger}}_{2,p_4}).
\end{equation}
   Thus ${\delta \mathcal{H} = \sum_{\{p_i \}} g_{\{ p_i \}} \mathcal{O}_{\{ p_i \}}}$ where ${g_{\{ p_i \}}= g_{p_1,p_2,p_3,p_4}}$ are coupling constants. Using Eq.~(\ref{eq:psi}) it is easy to see that this operator reads
\begin{multline}
\mathcal{O}_{p_1,p_2,p_3,p_4} \thicksim
  e^{i 2(p_1-p_2+p_3-p_4) \pi \frac{n}{2} x} \\
  \cdot e^{-i \sqrt{2} (p_1-p_2+p_3-p_4) \phi_\rho} \cdot e^{-i \sqrt{2} (p_1-p_2-p_3-p_4) \phi_\sigma}
  \end{multline}
We see that the $e^{i \theta_\mu}$'s, which signal particle creation, cancel. When the oscillating part is absent, as in the cases discussed below, the couplings $g_{\{ p_i \}}$ flow according to the RG equation ${\frac{dg_{\{ p_i \}} }{d \ell} = (2 - x_{\{ p_i \}}) dg_{\{ p_i \}}}$, with the scaling dimension
\begin{equation}
x_{\{ p_i \}} = \tfrac{(p_1-p_2+p_3-p_4)^2 K_\rho
+   (p_1-p_2-p_3+p_4)^2 K_\sigma}{2}   .
\end{equation}
Here, ${d \ell = d \ln \frac{\Lambda_0}{\Lambda}}$, with $\Lambda_0$ being the ultraviolet momentum cutoff. Such an operator becomes relevant for ${x_{\{ p_i \}}<2}$.
The leading operators are: (i) the spin density wave (SDW) operator obtained from Eq.~(\ref{psi4}) with ${(p_1-p_2) = -(p_3-p_4)=1}$ which is exactly marginal at ${K_\sigma=1}$ and nonoscillating; (ii) the level $\ell$ charge density wave (CDW) umklapp operator, obtained from Eq.~(\ref{psi4}) with ${(p_1-p_2)=(p_3-p_4)=\ell}$. This operator contains an oscillating exponent $e^{i 2 \pi \ell \frac{n}{2} x}$. On the lattice, $x$ takes integer values. Hence, such an operator is constant only  if ${\ell \frac{n}{2} \in \mathbb{N}}$.

We herein perturbatively include the inter-chain coupling $t_{\perp}$. Operators that do conserve the total but not the relative charge of the two chains are therefore allowed, leading to an additional list of operators ${\delta    \mathcal{H}_\perp}$, thus   ${\mathcal{H} = \mathcal{H}_\mathrm{LL} + \delta \mathcal{H}+ \delta \mathcal{H}_\perp}$. Using  Eq.~(\ref{eq:psi}), we may classify
the various inter-chain coupling operators. They originate from the operators ${(\psi^\dagger_{1,p'} \psi^{\phantom{\dagger}}_{2,p\phantom{'}})^q}$ with integer q. We first analyse the ${q=1}$ case
\begin{equation}
\mathcal{O}_{pp'} \thicksim  \psi^\dagger_{1,p'} \psi^{\phantom{\dagger}}_{2,p\phantom{'}},
\end{equation}
which takes the form of
\begin{align}
\label{cosgeneral}
\int dx  \cos [\theta_1 - \theta_2
+  (p-p')\pi n x +2(p' \phi_1-  p \phi_2 )+ \Phi x].
\end{align}
Thus, we include ${\delta \mathcal{H}_\perp = \sum_{p,p'} g_{pp'} \mathcal{O}_{pp'}}$ with $g_{pp'} \propto t_\perp$. Demanding that this operator commutes with itself at different points requires
\begin{equation}
p'=-p.
\end{equation}
These operators are oscillating except for
\begin{align}
\label{eq:genosc}
(p-p') \pi n=  \Phi,~~~(p \ge p').
\end{align}
The general operator $\mathcal{O}_{pp'}$ has a scaling dimension
\begin{equation}
x_{pp'} = \frac{1}{2} \left(\frac{1}{K_\sigma} +(p+p')^2 K_\sigma+  (p-p')^2 K_\rho \right).
\end{equation}
It corresponds to some of the following various interesting phases.

\emph{Meissner phase} - In the case of ${p=p'=0}$ (bosons) at zero flux the operator Eq.~(\ref{cosgeneral}) is relevant for ${K_\sigma<4}$. It is nonoscillating at zero flux,
while it becomes oscillating at nonzero flux. Yet, in a grand-canonical ensemble defined by a chemical potential $\mu$ rather than the density $n$, there is a finite range of nonzero flux where a gap opens, up to a commensurate-incommensurate transition~\cite{Orignac,Petrescu13}, even though the condition in  Eq.~(\ref{eq:genosc}) is not satisfied.

\emph{FQH state} - In the case of operators with ${p = - p' \ne 0}$, which matches Eq.~(\ref{cos}), the absence of oscillations occurs \emph{at finite flux} when the filling factor is given by ${\nu =\frac{ 1}{2p}}$, which is the inverse of an even(odd) integer for bosons(fermions).

\emph{Vortex lattice} - Another type of inter-chain coupling operators that could be included in ${\delta \mathcal{H}_\perp}$ originates from the special case of ${(\psi^\dagger_{1,0} \psi^{\phantom{\dagger}}_{2,0})^q }$, which becomes nonoscillating when the magnetic flux is commensurate~\cite{Orignac}, leading to a vortex lattice (VL) state.

The instabilities along with scaling dimensions and commensurability conditions are summarized in Table \ref{tb:operators}.

 \begin {table}
 \caption {Scaling dimensions and commensurability conditions of possible instabilities.
 }
\label{tb:operators}
\begin{center}
\begin{tabular}{l || c | c | c }
   instability & $\cos[\sqrt{2}(\dots)]$ &  dimension & commensurability   \\ \hline\hline
  SDW &  $2 \phi_\sigma$  & $\scriptstyle{2 K_\sigma}$ & -   \\ \hline
  CDW  &  $\ell \phi_\rho$  & $\scriptstyle{\frac{1}{2}\ell^2 K_\rho}$& $\ell \frac{n}{2} \in \mathbb{N}$  \\ \hline
   FQH &  $\theta_\sigma  + 2p \phi_\rho$  & $\scriptstyle{\frac{1}{2}\left(K_\sigma^{-1} +  (2 p)^2 K_\rho\right)}$ &  $\nu=\frac{\pi n}{\Phi} = \frac{1}{2 p }$ \\ \hline
   VL~\cite{Orignac} &
    $q  \theta_\sigma$ & $\scriptstyle{\frac{1}{2}K_\sigma^{-1}q^2}$ & $\frac{q \Phi}{2 \pi} \in \mathbb{N}$  (bosons) \\
\end{tabular}
\end{center}
\end{table}

\section{Solutions of the exactly solvable models}
\label{se:appendix1}
In this appendix we complete the derivation of the Luttinger charge-parameter $K_\rho$ for the two-leg ladder solvable model ${H_0+H_{int}}$ with hard core interaction of range $\xi$, presented in Sec.~\ref{se:specialmodel}. This is done using the methods of Ref.~\onlinecite{SelaGarst11}. We continue by generalizing this mapping to enable the treatment of a special class of finite $t_\perp$ models.

\subsection{Luttinger Parameter}
Since ${t_\perp=0}$, we start by mapping the two-leg ladder to a single chain of length ${L'=L-(N-1)\xi}$, and represent the low-energy space, which does \emph{not} contain pair of particles horizontally separated by $\xi$ sites or less, using free fermions. Their momenta in the reduced chain take values of
\begin{equation}
k_m = \frac{2 \pi m}{L'} = \frac{2 \pi m}{L(1-n \xi)},~~~m\in\mathbb{N}.
\end{equation}
Filling up the cosine dispersion, the ground state energy is
\begin{equation}
\label{eq:EGS}
E_\mathrm{GS}=-2 t \sum_{m= -\frac{1}{2}n L}^{\frac{1}{2}n L} \cos k_m.
\end{equation}
Thus, we can compute the inverse compressibility to be
\begin{equation}
\label{result1}
\frac{1}{L} \frac{\partial^2 E_\mathrm{GS}}{\partial n^2} = \frac{2 \pi}{(1- n \xi)^3} \sin \left( \frac{n  \pi}{1- n \xi}  \right).
\end{equation}
An additional thermodynamic quantity can be computed to determine the parameters of the low-energy theory. Consider an Aharonov-Bohm flux corresponding to a phase $\Phi_{L}$ in the original model when closed into a loop. It can be included in one of the links of the Hamiltonian. Alternatively, the same total phase can be included uniformly as a phase factor $e^{ i\frac{1}{L}\Phi_L}$ to every link. The fermions experience this uniform phase over $L'$ sites, and hence, their total Aharonov-Bohm phase is ${\Phi'_L =\Phi_L \frac{L'}{L}}$. In the presence of this phase, their momenta is quantized as
\begin{equation}
k_m = \frac{2 \pi m}{L'} +\frac{\Phi'_L}{L'}= \frac{2 \pi m}{L(1-n \xi)} +\frac{\Phi_L}{L},~~~m\in\mathbb{N}.
\end{equation}
The total energy is still computed with Eq.~(\ref{eq:EGS}).
Thus, we can calculate the phase stiffness,
\begin{equation}
\label{result2}
L \frac{\partial^2 E_\mathrm{GS}}{\partial \Phi_L^2} = \frac{2 (1- n \xi)}{\pi} \sin \left( \frac{n  \pi}{1- n \xi}  \right).
\end{equation}
Those two thermodynamic quantities are sufficient to compute the charge Luttinger parameter~\cite{Giamarchi}. The term ${(\nabla \phi_\rho)^2}$ is responsible for the compressibility, yielding the relation
\begin{equation}
\frac{1}{L} \frac{\partial^2 E_\mathrm{GS}}{\partial n^2}  = \frac{\pi v_\rho}{2 K_\rho}.
\end{equation}
The phase $\Phi_L$ corresponds to the same flux and hence to a vector potential ${A = \frac{1}{L}\Phi_L}$. The vector potential couples to the current $j_\rho$, which according to the continuity equation is given by
\begin{equation}
j_\rho = -i[\mathcal{H},-\tfrac{\sqrt{2}}{\pi}\phi_\rho]=\sqrt{2} v_\rho K_\rho \Pi_\rho.
\end{equation}
From the form of the coupling term ${\thicksim j_\rho A}$, one can immediately see that the second derivative of the energy with respect to the phase $\Phi_L$ is related to the coefficient of $j_\rho^2$ term in the Luttinger liquid Hamiltonian
\begin{equation}
L \frac{\partial^2 E_\mathrm{GS}}{\partial \Phi_L^2}  = \frac{2  v_\rho K_\rho}{\pi }.
\end{equation}
We hence obtain the Luttinger parameter
\begin{equation}
K_\rho =\frac{\pi}{2}\sqrt{\frac{L \frac{\partial^2 E_\mathrm{GS}}{\partial \Phi_L^2}  }{\frac{1}{L} \frac{\partial^2 E_\mathrm{GS}}{\partial n^2} }}.
\end{equation}
Therefore, using Eq.~(\ref{result1}) and (\ref{result2}), the low-energy physics of the model, Eq.~(\ref{intmodel}) and (\ref{interaction}), is described by a Luttinger liquid with
\begin{equation}
K_\rho = \frac{1}{2}(1- n \xi)^2.
\end{equation}
\subsection{Mapping to a $\nu=1$ phase of zero-range interacting fermions}
We may follow the same mapping from $L$ rungs to ${L' = L-(N-1)\xi}$ rungs in the limit ${U \to \infty}$ at finite $t_\perp$, and map the two-leg ladder to a reduced two-leg ladder. We restrict our attention to fermions, with the treatment of bosons generalized below. The part ${H_0+H_{\mathrm{int}}}$ takes the shape of a ${\xi' = 0}$ Hamiltonian. Since site $j$ in the new lattice correspond to location ${j+\xi(\sum_{j'=1}^{j-1} n_{j'})}$ in the original lattice, the inter-chain coupling becomes
\begin{equation}
\label{intHperp}
H_{\perp}' = - t_\perp \sum_j \left(c^\dagger_{j,1} c^{\phantom{\dagger}}_{j,2} e^{i \Phi [j+\xi(\sum_{j'=1}^{j-1} n_{j'})]}+ \mathrm{h.c.}\right),
\end{equation}
which is nonlocal. However, the nonlocality disappears for a special value of the flux
\begin{equation}
\Phi = \frac{2\pi m}{\xi},
\end{equation}
where $m$ is a nonnegative integer. Thus, the new particles are subject to the same value of flux ${\Phi'=\Phi}$.
For a given filling ${\nu}$, the density of the original particles is
\begin{equation}
n = \nu \frac{\Phi}{\pi} =\nu\frac{2 m}{\xi}.
 \end{equation}
In the thermodynamic limit,
$L' = L - N \xi$, and therefore, ${\frac{L'}{L} = \frac{\nu}{\nu'} = 1-n\xi = 1 - \nu 2 m }$.  The density of the new particles is ${n'=\frac{N}{L'}=\frac{1}{n^{-1}-\xi}}$, and
the new filling factor is thus
\begin{equation}
\nu'=\frac{n'\pi}{\Phi'}=\frac{1}{\nu^{-1} - 2 m}.
\end{equation}
Since ${\xi=0}$ in the new system, the Luttinger parameter is ${K_\rho'=\frac{1}{2}}$ and the scaling dimension $x$ may be expressed in either the new system or the original one,
\begin{equation}
x=\frac{1}{2} \left(1+ (\nu')^{-2} \cdot\frac{1}{2}\right) = \frac{1}{2} \left(1+ \nu^{-2}\cdot\frac{1}{2}(1- n\xi)^2\right).
\end{equation}
As the density should not exceed $\frac{1}{\xi+1}$, the integer $m$ is a constrained by ${m < \frac{1}{2 \nu} \frac{\xi}{\xi+1}}$. One interesting case where the constraint is satisfied is
\begin{equation}
\label{EXACTmpa}
\begin{array}{llll}
\nu\phantom{'}=\frac{1}{2m+1},&\xi\phantom{'} = 2 m +1,&\Phi\phantom{'} =\frac{2 \pi m }{2m +1},&n = \frac{2m}{(2m+1)^2}, \\
\nu'=1,&\xi'=0,&\Phi'=\frac{2 \pi m }{2m +1},& n' = \frac{2m}{2m+1}.
\end{array}
\end{equation}
For example, ${m=1}$ corresponds to ${\nu=\frac{1}{3}}$, ${\xi=1}$, ${
 \Phi = \frac{2 \pi}{3}}$, ${n=\frac{2}{9}}$, ${n'=\frac{2}{3}}$.
This example demonstrates that the ${\nu'=1}$ gap for the ${\xi'=0}$ model is equivalent to ${\nu=\frac{1}{3}}$ of the original particles. It is always relevant as ${x = \frac{3}{4}}$, showing that there should be gap opening at small $t_\perp$.

\hfill

Starting from bosons, one can reach similar mapping by including a Jordan-Wigner transformation~\cite{Giamarchi}. This leads to Eq.~(\ref{EXACTmpa}) with half integer $m$. For example, ${m=\frac{1}{2}}$ corresponds to ${\nu=\frac{1}{2}}$, ${\xi=2}$, ${
 \Phi = \frac{\pi}{2}}$, ${n=\frac{1}{4}}$, ${n'=\frac{1}{2}}$. Thus, it constitutes a mapping of the bosonic ${\nu=\frac{1}{2}}$ state to the fermionic ${\nu'=1}$ state.

\hfill

The choice of parameters discussed here illustrates an important point. For weak interactions, the coefficient of the cosine operator driving the FQH instability is perturbatively small, behaving as a high power of the interactions~\cite{OregSelaStern}. However, Eq.~(\ref{intHperp}) implies that for very strong interactions, the coefficient of the cosine operator becomes simply $t_\perp$, as in the non-interacting fermionic case. Indeed, this is observed by substituting the bosonized expression for the density,
\begin{equation}
\sum_{j'=1}^{j-1} n_{j'} =n\cdot x-\frac{\sqrt{2}}{\pi}\phi_\rho (x),
\end{equation}
into Eq.~(\ref{intHperp}).
Upon bosonizing $c^\dagger_{j,1} c^{\phantom{\dagger}}_{j,2}$, the cosine operator Eq.~(\ref{cos}) is recovered with $g_p=t_\perp$ for $\Phi \xi= 2 \pi m$.

Thus, while in the weakly interacting case, the cosine interaction Eq.~(\ref{cos}) is generated by RG, we see via this explicit case study, that the FQH coupling constant, $g_p$, is non-perturbative in the strongly interacting regime.


\begin{thebibliography}{49}
\expandafter\ifx\csname natexlab\endcsname\relax\def\natexlab#1{#1}\fi
\expandafter\ifx\csname bibnamefont\endcsname\relax
  \def\bibnamefont#1{#1}\fi
\expandafter\ifx\csname bibfnamefont\endcsname\relax
  \def\bibfnamefont#1{#1}\fi
\expandafter\ifx\csname citenamefont\endcsname\relax
  \def\citenamefont#1{#1}\fi
\expandafter\ifx\csname url\endcsname\relax
  \def\url#1{\texttt{#1}}\fi
\expandafter\ifx\csname urlprefix\endcsname\relax\def\urlprefix{URL }\fi
\providecommand{\bibinfo}[2]{#2}
\providecommand{\eprint}[2][]{\url{#2}}

\bibitem[{\citenamefont{Atala et~al.}(2014)\citenamefont{Atala, Aidelsburger,
  Lohse, Barreiro, Paredes, and Bloch}}]{Bloch14}
\bibinfo{author}{\bibfnamefont{M.}~\bibnamefont{Atala}},
  \bibinfo{author}{\bibfnamefont{M.}~\bibnamefont{Aidelsburger}},
  \bibinfo{author}{\bibfnamefont{M.}~\bibnamefont{Lohse}},
  \bibinfo{author}{\bibfnamefont{J.~T.} \bibnamefont{Barreiro}},
  \bibinfo{author}{\bibfnamefont{B.}~\bibnamefont{Paredes}}, \bibnamefont{and}
  \bibinfo{author}{\bibfnamefont{I.}~\bibnamefont{Bloch}},
  \bibinfo{journal}{Nature Physics} \textbf{\bibinfo{volume}{10}},
  \bibinfo{pages}{588} (\bibinfo{year}{2014}).

\bibitem[{\citenamefont{Aidelsburger et~al.}(2011)\citenamefont{Aidelsburger,
  Atala, Nascimb\`ene, Trotzky, Chen, and Bloch}}]{Aidelsburger}
\bibinfo{author}{\bibfnamefont{M.}~\bibnamefont{Aidelsburger}},
  \bibinfo{author}{\bibfnamefont{M.}~\bibnamefont{Atala}},
  \bibinfo{author}{\bibfnamefont{S.}~\bibnamefont{Nascimb\`ene}},
  \bibinfo{author}{\bibfnamefont{S.}~\bibnamefont{Trotzky}},
  \bibinfo{author}{\bibfnamefont{Y.-A.} \bibnamefont{Chen}}, \bibnamefont{and}
  \bibinfo{author}{\bibfnamefont{I.}~\bibnamefont{Bloch}},
  \bibinfo{journal}{Phys. Rev. Lett.} \textbf{\bibinfo{volume}{107}},
  \bibinfo{pages}{255301} (\bibinfo{year}{2011}).

\bibitem[{\citenamefont{Goldman et~al.}(2014)\citenamefont{Goldman,
  Juzeli{\=u}nas, {\"O}hberg, and Spielman}}]{Goldman13}
\bibinfo{author}{\bibfnamefont{N.}~\bibnamefont{Goldman}},
  \bibinfo{author}{\bibfnamefont{G.}~\bibnamefont{Juzeli{\=u}nas}},
  \bibinfo{author}{\bibfnamefont{P.}~\bibnamefont{{\"O}hberg}},
  \bibnamefont{and} \bibinfo{author}{\bibfnamefont{I.~B.}
  \bibnamefont{Spielman}}, \bibinfo{journal}{Reports on Progress in Physics}
  \textbf{\bibinfo{volume}{77}}, \bibinfo{pages}{126401}
  (\bibinfo{year}{2014}).

\bibitem[{\citenamefont{Jaksch and Zoller}(2003)}]{Zoller03}
\bibinfo{author}{\bibfnamefont{D.}~\bibnamefont{Jaksch}} \bibnamefont{and}
  \bibinfo{author}{\bibfnamefont{P.}~\bibnamefont{Zoller}},
  \bibinfo{journal}{New Journal of Physics} \textbf{\bibinfo{volume}{5}},
  \bibinfo{pages}{56} (\bibinfo{year}{2003}).

\bibitem[{\citenamefont{Gerbier and Dalibard}(2010)}]{Gerbier10}
\bibinfo{author}{\bibfnamefont{F.}~\bibnamefont{Gerbier}} \bibnamefont{and}
  \bibinfo{author}{\bibfnamefont{J.}~\bibnamefont{Dalibard}},
  \bibinfo{journal}{New Journal of Physics} \textbf{\bibinfo{volume}{12}},
  \bibinfo{pages}{033007} (\bibinfo{year}{2010}).

\bibitem[{\citenamefont{Kolovsky}(2011)}]{Kolovsky}
\bibinfo{author}{\bibfnamefont{A.~R.} \bibnamefont{Kolovsky}},
  \bibinfo{journal}{EPL (Europhysics Letters)} \textbf{\bibinfo{volume}{93}},
  \bibinfo{pages}{20003} (\bibinfo{year}{2011}).

\bibitem[{\citenamefont{H\"ugel and Paredes}(2014)}]{hugel}
\bibinfo{author}{\bibfnamefont{D.}~\bibnamefont{H\"ugel}} \bibnamefont{and}
  \bibinfo{author}{\bibfnamefont{B.}~\bibnamefont{Paredes}},
  \bibinfo{journal}{Phys. Rev. A} \textbf{\bibinfo{volume}{89}},
  \bibinfo{pages}{023619} (\bibinfo{year}{2014}).

\bibitem[{kan()}]{kane2002}
\bibinfo{note}{C. L. Kane, R. Mukhopadhyay, and T. C. Lubensky, Phys. Rev.
  Lett. \textbf{88}, 036401 (2002); J. C. Y. Teo and C. L. Kane, Phys. Rev. B
  \textbf{89}, 085101 (2014).}

\bibitem[{\citenamefont{Petrescu and Le~Hur}(2015)}]{Petrescu}
\bibinfo{author}{\bibfnamefont{A.}~\bibnamefont{Petrescu}} \bibnamefont{and}
  \bibinfo{author}{\bibfnamefont{K.}~\bibnamefont{Le~Hur}},
  \bibinfo{journal}{Phys. Rev. B} \textbf{\bibinfo{volume}{91}},
  \bibinfo{pages}{054520} (\bibinfo{year}{2015}).

\bibitem[{\citenamefont{Grusdt and H\"oning}(2014)}]{Grusdt}
\bibinfo{author}{\bibfnamefont{F.}~\bibnamefont{Grusdt}} \bibnamefont{and}
  \bibinfo{author}{\bibfnamefont{M.}~\bibnamefont{H\"oning}},
  \bibinfo{journal}{Phys. Rev. A} \textbf{\bibinfo{volume}{90}},
  \bibinfo{pages}{053623} (\bibinfo{year}{2014}).

\bibitem[{\citenamefont{Oreg et~al.}(2010)\citenamefont{Oreg, Refael, and von
  Oppen}}]{oreg}
\bibinfo{author}{\bibfnamefont{Y.}~\bibnamefont{Oreg}},
  \bibinfo{author}{\bibfnamefont{G.}~\bibnamefont{Refael}}, \bibnamefont{and}
  \bibinfo{author}{\bibfnamefont{F.}~\bibnamefont{von Oppen}},
  \bibinfo{journal}{Phys. Rev. Lett.} \textbf{\bibinfo{volume}{105}},
  \bibinfo{pages}{177002} (\bibinfo{year}{2010}).

\bibitem[{\citenamefont{Lutchyn et~al.}(2010)\citenamefont{Lutchyn, Sau, and
  Das~Sarma}}]{lutchin}
\bibinfo{author}{\bibfnamefont{R.~M.} \bibnamefont{Lutchyn}},
  \bibinfo{author}{\bibfnamefont{J.~D.} \bibnamefont{Sau}}, \bibnamefont{and}
  \bibinfo{author}{\bibfnamefont{S.}~\bibnamefont{Das~Sarma}},
  \bibinfo{journal}{Phys. Rev. Lett.} \textbf{\bibinfo{volume}{105}},
  \bibinfo{pages}{077001} (\bibinfo{year}{2010}).

\bibitem[{\citenamefont{Oreg et~al.}(2014)\citenamefont{Oreg, Sela, and
  Stern}}]{OregSelaStern}
\bibinfo{author}{\bibfnamefont{Y.}~\bibnamefont{Oreg}},
  \bibinfo{author}{\bibfnamefont{E.}~\bibnamefont{Sela}}, \bibnamefont{and}
  \bibinfo{author}{\bibfnamefont{A.}~\bibnamefont{Stern}},
  \bibinfo{journal}{Phys. Rev. B} \textbf{\bibinfo{volume}{89}},
  \bibinfo{pages}{115402} (\bibinfo{year}{2014}).

\bibitem[{Lin()}]{Lin}
\bibinfo{note}{Y.-J. Lin, K. Jim\'{e}nez-Garc\'{\i}a, and I. B. Spielman,
  Nature \textbf{471}, 83 (2011).}

\bibitem[{\citenamefont{Cheuk et~al.}(2012)\citenamefont{Cheuk, Sommer,
  Hadzibabic, Yefsah, Bakr, and Zwierlein}}]{Cheuk}
\bibinfo{author}{\bibfnamefont{L.~W.} \bibnamefont{Cheuk}},
  \bibinfo{author}{\bibfnamefont{A.~T.} \bibnamefont{Sommer}},
  \bibinfo{author}{\bibfnamefont{Z.}~\bibnamefont{Hadzibabic}},
  \bibinfo{author}{\bibfnamefont{T.}~\bibnamefont{Yefsah}},
  \bibinfo{author}{\bibfnamefont{W.~S.} \bibnamefont{Bakr}}, \bibnamefont{and}
  \bibinfo{author}{\bibfnamefont{M.~W.} \bibnamefont{Zwierlein}},
  \bibinfo{journal}{Phys. Rev. Lett.} \textbf{\bibinfo{volume}{109}},
  \bibinfo{pages}{095302} (\bibinfo{year}{2012}).

\bibitem[{\citenamefont{Wang et~al.}(2012)\citenamefont{Wang, Yu, Fu, Miao,
  Huang, Chai, Zhai, and Zhang}}]{Wang}
\bibinfo{author}{\bibfnamefont{P.}~\bibnamefont{Wang}},
  \bibinfo{author}{\bibfnamefont{Z.-Q.} \bibnamefont{Yu}},
  \bibinfo{author}{\bibfnamefont{Z.}~\bibnamefont{Fu}},
  \bibinfo{author}{\bibfnamefont{J.}~\bibnamefont{Miao}},
  \bibinfo{author}{\bibfnamefont{L.}~\bibnamefont{Huang}},
  \bibinfo{author}{\bibfnamefont{S.}~\bibnamefont{Chai}},
  \bibinfo{author}{\bibfnamefont{H.}~\bibnamefont{Zhai}}, \bibnamefont{and}
  \bibinfo{author}{\bibfnamefont{J.}~\bibnamefont{Zhang}},
  \bibinfo{journal}{Phys. Rev. Lett.} \textbf{\bibinfo{volume}{109}},
  \bibinfo{pages}{095301} (\bibinfo{year}{2012}).

\bibitem[{\citenamefont{Cui et~al.}(2013)\citenamefont{Cui, Lian, Ho, Lev, and
  Zhai}}]{Cui}
\bibinfo{author}{\bibfnamefont{X.}~\bibnamefont{Cui}},
  \bibinfo{author}{\bibfnamefont{B.}~\bibnamefont{Lian}},
  \bibinfo{author}{\bibfnamefont{T.-L.} \bibnamefont{Ho}},
  \bibinfo{author}{\bibfnamefont{B.~L.} \bibnamefont{Lev}}, \bibnamefont{and}
  \bibinfo{author}{\bibfnamefont{H.}~\bibnamefont{Zhai}},
  \bibinfo{journal}{Phys. Rev. A} \textbf{\bibinfo{volume}{88}},
  \bibinfo{pages}{011601} (\bibinfo{year}{2013}).

\bibitem[{\citenamefont{{Mancini} et~al.}(2015)\citenamefont{{Mancini},
  {Pagano}, {Cappellini}, {Livi}, {Rider}, {Catani}, {Sias}, {Zoller},
  {Inguscio}, {Dalmonte} et~al.}}]{Mancini}
\bibinfo{author}{\bibfnamefont{M.}~\bibnamefont{{Mancini}}},
  \bibinfo{author}{\bibfnamefont{G.}~\bibnamefont{{Pagano}}},
  \bibinfo{author}{\bibfnamefont{G.}~\bibnamefont{{Cappellini}}},
  \bibinfo{author}{\bibfnamefont{L.}~\bibnamefont{{Livi}}},
  \bibinfo{author}{\bibfnamefont{M.}~\bibnamefont{{Rider}}},
  \bibinfo{author}{\bibfnamefont{J.}~\bibnamefont{{Catani}}},
  \bibinfo{author}{\bibfnamefont{C.}~\bibnamefont{{Sias}}},
  \bibinfo{author}{\bibfnamefont{P.}~\bibnamefont{{Zoller}}},
  \bibinfo{author}{\bibfnamefont{M.}~\bibnamefont{{Inguscio}}},
  \bibinfo{author}{\bibfnamefont{M.}~\bibnamefont{{Dalmonte}}},
  \bibnamefont{et~al.}, \bibinfo{journal}{ArXiv e-prints}
  (\bibinfo{year}{2015}), \eprint{1502.02495}.

\bibitem[{\citenamefont{{Stuhl} et~al.}(2015)\citenamefont{{Stuhl}, {Lu},
  {Aycock}, {Genkina}, and {Spielman}}}]{Stuhl}
\bibinfo{author}{\bibfnamefont{B.~K.} \bibnamefont{{Stuhl}}},
  \bibinfo{author}{\bibfnamefont{H.-I.} \bibnamefont{{Lu}}},
  \bibinfo{author}{\bibfnamefont{L.~M.} \bibnamefont{{Aycock}}},
  \bibinfo{author}{\bibfnamefont{D.}~\bibnamefont{{Genkina}}},
  \bibnamefont{and} \bibinfo{author}{\bibfnamefont{I.~B.}
  \bibnamefont{{Spielman}}}, \bibinfo{journal}{ArXiv e-prints}
  (\bibinfo{year}{2015}), \eprint{1502.02496}.

\bibitem[{\citenamefont{{Barbarino} et~al.}(2015)\citenamefont{{Barbarino},
  {Taddia}, {Rossini}, {Mazza}, and {Fazio}}}]{Fazio}
\bibinfo{author}{\bibfnamefont{S.}~\bibnamefont{{Barbarino}}},
  \bibinfo{author}{\bibfnamefont{L.}~\bibnamefont{{Taddia}}},
  \bibinfo{author}{\bibfnamefont{D.}~\bibnamefont{{Rossini}}},
  \bibinfo{author}{\bibfnamefont{L.}~\bibnamefont{{Mazza}}}, \bibnamefont{and}
  \bibinfo{author}{\bibfnamefont{R.}~\bibnamefont{{Fazio}}},
  \bibinfo{journal}{ArXiv e-prints}  (\bibinfo{year}{2015}),
  \eprint{1504.00164}.

\bibitem[{\citenamefont{{Zeng} et~al.}(2015)\citenamefont{{Zeng}, {Wang}, and
  {Zhai}}}]{Zeng}
\bibinfo{author}{\bibfnamefont{T.-S.} \bibnamefont{{Zeng}}},
  \bibinfo{author}{\bibfnamefont{C.}~\bibnamefont{{Wang}}}, \bibnamefont{and}
  \bibinfo{author}{\bibfnamefont{H.}~\bibnamefont{{Zhai}}},
  \bibinfo{journal}{ArXiv e-prints}  (\bibinfo{year}{2015}),
  \eprint{1504.02263}.

\bibitem[{\citenamefont{{Yan} et~al.}(2015)\citenamefont{{Yan}, {Wan}, and
  {Wang}}}]{Majoranacoldatoms}
\bibinfo{author}{\bibfnamefont{Z.}~\bibnamefont{{Yan}}},
  \bibinfo{author}{\bibfnamefont{S.}~\bibnamefont{{Wan}}}, \bibnamefont{and}
  \bibinfo{author}{\bibfnamefont{Z.}~\bibnamefont{{Wang}}},
  \bibinfo{journal}{ArXiv e-prints}  (\bibinfo{year}{2015}),
  \eprint{1504.03223}.

\bibitem[{\citenamefont{Meng et~al.}(2014)\citenamefont{Meng, Fritz, Schuricht,
  and Loss}}]{Meng14}
\bibinfo{author}{\bibfnamefont{T.}~\bibnamefont{Meng}},
  \bibinfo{author}{\bibfnamefont{L.}~\bibnamefont{Fritz}},
  \bibinfo{author}{\bibfnamefont{D.}~\bibnamefont{Schuricht}},
  \bibnamefont{and} \bibinfo{author}{\bibfnamefont{D.}~\bibnamefont{Loss}},
  \bibinfo{journal}{Phys. Rev. B} \textbf{\bibinfo{volume}{89}},
  \bibinfo{pages}{045111} (\bibinfo{year}{2014}).

\bibitem[{\citenamefont{Cornfeld et~al.}(2015)\citenamefont{Cornfeld, Neder,
  and Sela}}]{Cornfeld15}
\bibinfo{author}{\bibfnamefont{E.}~\bibnamefont{Cornfeld}},
  \bibinfo{author}{\bibfnamefont{I.}~\bibnamefont{Neder}}, \bibnamefont{and}
  \bibinfo{author}{\bibfnamefont{E.}~\bibnamefont{Sela}},
  \bibinfo{journal}{Phys. Rev. B} \textbf{\bibinfo{volume}{91}},
  \bibinfo{pages}{115427} (\bibinfo{year}{2015}).

\bibitem[{\citenamefont{Orignac and Giamarchi}(2001)}]{Orignac}
\bibinfo{author}{\bibfnamefont{E.}~\bibnamefont{Orignac}} \bibnamefont{and}
  \bibinfo{author}{\bibfnamefont{T.}~\bibnamefont{Giamarchi}},
  \bibinfo{journal}{Phys. Rev. B} \textbf{\bibinfo{volume}{64}},
  \bibinfo{pages}{144515} (\bibinfo{year}{2001}).

\bibitem[{\citenamefont{Petrescu and Le~Hur}(2013)}]{Petrescu13}
\bibinfo{author}{\bibfnamefont{A.}~\bibnamefont{Petrescu}} \bibnamefont{and}
  \bibinfo{author}{\bibfnamefont{K.}~\bibnamefont{Le~Hur}},
  \bibinfo{journal}{Phys. Rev. Lett.} \textbf{\bibinfo{volume}{111}},
  \bibinfo{pages}{150601} (\bibinfo{year}{2013}).

\bibitem[{\citenamefont{Piraud et~al.}(2015)\citenamefont{Piraud,
  Heidrich-Meisner, McCulloch, Greschner, Vekua, and Schollw\"ock}}]{Piraud}
\bibinfo{author}{\bibfnamefont{M.}~\bibnamefont{Piraud}},
  \bibinfo{author}{\bibfnamefont{F.}~\bibnamefont{Heidrich-Meisner}},
  \bibinfo{author}{\bibfnamefont{I.~P.} \bibnamefont{McCulloch}},
  \bibinfo{author}{\bibfnamefont{S.}~\bibnamefont{Greschner}},
  \bibinfo{author}{\bibfnamefont{T.}~\bibnamefont{Vekua}}, \bibnamefont{and}
  \bibinfo{author}{\bibfnamefont{U.}~\bibnamefont{Schollw\"ock}},
  \bibinfo{journal}{Phys. Rev. B} \textbf{\bibinfo{volume}{91}},
  \bibinfo{pages}{140406} (\bibinfo{year}{2015}).

\bibitem[{Mue()}]{Mueller14}
\bibinfo{note}{E. Mueller, Nature Physics \textbf{10}, 554–555 (2014).}

\bibitem[{Yi0()}]{Yi06}
\bibinfo{note}{W. Yi, S. Wei and Z. Guang-Hui, Chinese Phys. Lett. \textbf{23}
  3065 (2006).}

\bibitem[{Gia()}]{Giamarchi}
\bibinfo{note}{T. Giamarchi, {\it Quantum Physics in One Dimension} (Oxford
  University Press, New York, 2004).}

\bibitem[{\citenamefont{Sela and Pereira}(2011)}]{SelaPereiraORBITAL}
\bibinfo{author}{\bibfnamefont{E.}~\bibnamefont{Sela}} \bibnamefont{and}
  \bibinfo{author}{\bibfnamefont{R.~G.} \bibnamefont{Pereira}},
  \bibinfo{journal}{Phys. Rev. B} \textbf{\bibinfo{volume}{84}},
  \bibinfo{pages}{014407} (\bibinfo{year}{2011}).

\bibitem[{\citenamefont{Sela et~al.}(2011)\citenamefont{Sela, Punk, and
  Garst}}]{SelaGarst11}
\bibinfo{author}{\bibfnamefont{E.}~\bibnamefont{Sela}},
  \bibinfo{author}{\bibfnamefont{M.}~\bibnamefont{Punk}}, \bibnamefont{and}
  \bibinfo{author}{\bibfnamefont{M.}~\bibnamefont{Garst}},
  \bibinfo{journal}{Phys. Rev. B} \textbf{\bibinfo{volume}{84}},
  \bibinfo{pages}{085434} (\bibinfo{year}{2011}).

\bibitem[{\citenamefont{Gra\ss{} et~al.}(2015)\citenamefont{Gra\ss{}, Muschik,
  Celi, Chhajlany, and Lewenstein}}]{Grass}
\bibinfo{author}{\bibfnamefont{T.}~\bibnamefont{Gra\ss{}}},
  \bibinfo{author}{\bibfnamefont{C.}~\bibnamefont{Muschik}},
  \bibinfo{author}{\bibfnamefont{A.}~\bibnamefont{Celi}},
  \bibinfo{author}{\bibfnamefont{R.~W.} \bibnamefont{Chhajlany}},
  \bibnamefont{and}
  \bibinfo{author}{\bibfnamefont{M.}~\bibnamefont{Lewenstein}},
  \bibinfo{journal}{Phys. Rev. A} \textbf{\bibinfo{volume}{91}},
  \bibinfo{pages}{063612} (\bibinfo{year}{2015}).

\bibitem[{\citenamefont{Narozhny et~al.}(2005)\citenamefont{Narozhny, Carr, and
  Nersesyan}}]{Narozhny}
\bibinfo{author}{\bibfnamefont{B.~N.} \bibnamefont{Narozhny}},
  \bibinfo{author}{\bibfnamefont{S.~T.} \bibnamefont{Carr}}, \bibnamefont{and}
  \bibinfo{author}{\bibfnamefont{A.~A.} \bibnamefont{Nersesyan}},
  \bibinfo{journal}{Phys. Rev. B} \textbf{\bibinfo{volume}{71}},
  \bibinfo{pages}{161101} (\bibinfo{year}{2005}).

\bibitem[{\citenamefont{Carr et~al.}(2006)\citenamefont{Carr, Narozhny, and
  Nersesyan}}]{Carr}
\bibinfo{author}{\bibfnamefont{S.~T.} \bibnamefont{Carr}},
  \bibinfo{author}{\bibfnamefont{B.~N.} \bibnamefont{Narozhny}},
  \bibnamefont{and} \bibinfo{author}{\bibfnamefont{A.~A.}
  \bibnamefont{Nersesyan}}, \bibinfo{journal}{Phys. Rev. B}
  \textbf{\bibinfo{volume}{73}}, \bibinfo{pages}{195114}
  (\bibinfo{year}{2006}).

\bibitem[{\citenamefont{Simon and Affleck}(2001)}]{Simon01}
\bibinfo{author}{\bibfnamefont{P.}~\bibnamefont{Simon}} \bibnamefont{and}
  \bibinfo{author}{\bibfnamefont{I.}~\bibnamefont{Affleck}},
  \bibinfo{journal}{Phys. Rev. B} \textbf{\bibinfo{volume}{64}},
  \bibinfo{pages}{085308} (\bibinfo{year}{2001}).

\bibitem[{wen()}]{wen_book}
\bibinfo{note}{X.-G. Wen, {\it Quantum Field Theory of Many-body Systems}
  (Oxford University Press, New York, 2007).}

\bibitem[{Eza()}]{Ezawa}
\bibinfo{note}{Zyun. F. Ezawa, {\it Quantum Hall Effects} (World Scientific,
  Singapore, 2013) (page 317).}

\bibitem[{\citenamefont{Imambekov et~al.}(2012)\citenamefont{Imambekov,
  Schmidt, and Glazman}}]{Imambekov}
\bibinfo{author}{\bibfnamefont{A.}~\bibnamefont{Imambekov}},
  \bibinfo{author}{\bibfnamefont{T.~L.} \bibnamefont{Schmidt}},
  \bibnamefont{and} \bibinfo{author}{\bibfnamefont{L.~I.}
  \bibnamefont{Glazman}}, \bibinfo{journal}{Rev. Mod. Phys.}
  \textbf{\bibinfo{volume}{84}}, \bibinfo{pages}{1253} (\bibinfo{year}{2012}).

\bibitem[{\citenamefont{Pereira and Sela}(2010)}]{SelaPereira}
\bibinfo{author}{\bibfnamefont{R.~G.} \bibnamefont{Pereira}} \bibnamefont{and}
  \bibinfo{author}{\bibfnamefont{E.}~\bibnamefont{Sela}},
  \bibinfo{journal}{Phys. Rev. B} \textbf{\bibinfo{volume}{82}},
  \bibinfo{pages}{115324} (\bibinfo{year}{2010}).

\bibitem[{\citenamefont{Dhar et~al.}(2013)\citenamefont{Dhar, Mishra, Maji,
  Pai, Mukerjee, and Paramekanti}}]{Dhar}
\bibinfo{author}{\bibfnamefont{A.}~\bibnamefont{Dhar}},
  \bibinfo{author}{\bibfnamefont{T.}~\bibnamefont{Mishra}},
  \bibinfo{author}{\bibfnamefont{M.}~\bibnamefont{Maji}},
  \bibinfo{author}{\bibfnamefont{R.~V.} \bibnamefont{Pai}},
  \bibinfo{author}{\bibfnamefont{S.}~\bibnamefont{Mukerjee}}, \bibnamefont{and}
  \bibinfo{author}{\bibfnamefont{A.}~\bibnamefont{Paramekanti}},
  \bibinfo{journal}{Phys. Rev. B} \textbf{\bibinfo{volume}{87}},
  \bibinfo{pages}{174501} (\bibinfo{year}{2013}).

\bibitem[{\citenamefont{Wei and Mueller}(2014)}]{Wei}
\bibinfo{author}{\bibfnamefont{R.}~\bibnamefont{Wei}} \bibnamefont{and}
  \bibinfo{author}{\bibfnamefont{E.~J.} \bibnamefont{Mueller}},
  \bibinfo{journal}{Phys. Rev. A} \textbf{\bibinfo{volume}{89}},
  \bibinfo{pages}{063617} (\bibinfo{year}{2014}).

\bibitem[{\citenamefont{Kele\ifmmode~\mbox{\c{s}}\else \c{s}\fi{} and
  Oktel}(2015)}]{Ahmet}
\bibinfo{author}{\bibfnamefont{A.}~\bibnamefont{Kele\ifmmode~\mbox{\c{s}}\else
  \c{s}\fi{}}} \bibnamefont{and} \bibinfo{author}{\bibfnamefont{M.~O.}
  \bibnamefont{Oktel}}, \bibinfo{journal}{Phys. Rev. A}
  \textbf{\bibinfo{volume}{91}}, \bibinfo{pages}{013629}
  (\bibinfo{year}{2015}).

\bibitem[{\citenamefont{{Natu}}(2015)}]{Natu}
\bibinfo{author}{\bibfnamefont{S.~S.} \bibnamefont{{Natu}}},
  \bibinfo{journal}{ArXiv e-prints}  (\bibinfo{year}{2015}),
  \eprint{1506.04346}.

\bibitem[{Luk()}]{LukinDipoleBlockade}
\bibinfo{note}{M. D. Lukin, M. Fleischhauer, R. Cote, L. M. Duan, D. Jaksch, J.
  I. Cirac, and P. Zoller, Phys. Rev. Lett. \textbf{87}, 037901 (2001); D.
  Jaksch, J. I. Cirac, P. Zoller, S. L. Rolston, R. C\^ot\'e, and M. D. Lukin,
  \emph{ibid}. \textbf{85}, 2208 (2000).}

\bibitem[{Urb()}]{UrbanDipoleBlockade}
\bibinfo{note}{E. Urban, T. A. Johnson, T. Henage, L. Isenhower, D. D. Yavuz,
  T. G. Walker, and M. Saffman, Nat. Phys. \textbf{5}, 110 (2009); A.
  Ga{\"e}tan, Y. Miroshnychenko, T. Wilk, A. Chotia, M. Viteau, D. Comparat, P.
  Pillet, A. Browaeys, and P. Grangier, \emph{ibid.} \textbf{5}, 115 (2009).}

\bibitem[{\citenamefont{Weimer and B\"uchler}(2010)}]{Weimer10}
\bibinfo{author}{\bibfnamefont{H.}~\bibnamefont{Weimer}} \bibnamefont{and}
  \bibinfo{author}{\bibfnamefont{H.~P.} \bibnamefont{B\"uchler}},
  \bibinfo{journal}{Phys. Rev. Lett.} \textbf{\bibinfo{volume}{105}},
  \bibinfo{pages}{230403} (\bibinfo{year}{2010}).

\bibitem[{\citenamefont{Schauß et~al.}(2015)\citenamefont{Schauß, Zeiher,
  Fukuhara, Hild, Cheneau, Macrì, Pohl, Bloch, and Gross}}]{Schauss}
\bibinfo{author}{\bibfnamefont{P.}~\bibnamefont{Schauß}},
  \bibinfo{author}{\bibfnamefont{J.}~\bibnamefont{Zeiher}},
  \bibinfo{author}{\bibfnamefont{T.}~\bibnamefont{Fukuhara}},
  \bibinfo{author}{\bibfnamefont{S.}~\bibnamefont{Hild}},
  \bibinfo{author}{\bibfnamefont{M.}~\bibnamefont{Cheneau}},
  \bibinfo{author}{\bibfnamefont{T.}~\bibnamefont{Macrì}},
  \bibinfo{author}{\bibfnamefont{T.}~\bibnamefont{Pohl}},
  \bibinfo{author}{\bibfnamefont{I.}~\bibnamefont{Bloch}}, \bibnamefont{and}
  \bibinfo{author}{\bibfnamefont{C.}~\bibnamefont{Gross}},
  \bibinfo{journal}{Science} \textbf{\bibinfo{volume}{347}},
  \bibinfo{pages}{1455} (\bibinfo{year}{2015}).

\bibitem[{\citenamefont{Dalmonte et~al.}(2010)\citenamefont{Dalmonte, Pupillo,
  and Zoller}}]{Dalmonte10}
\bibinfo{author}{\bibfnamefont{M.}~\bibnamefont{Dalmonte}},
  \bibinfo{author}{\bibfnamefont{G.}~\bibnamefont{Pupillo}}, \bibnamefont{and}
  \bibinfo{author}{\bibfnamefont{P.}~\bibnamefont{Zoller}},
  \bibinfo{journal}{Phys. Rev. Lett.} \textbf{\bibinfo{volume}{105}},
  \bibinfo{pages}{140401} (\bibinfo{year}{2010}).

\end{thebibliography}
\end{document}